\documentclass[a4paper,fleqn,usenatbib]{mnras}
\usepackage[T1]{fontenc}
\usepackage{ae,aecompl}
\usepackage[dvips]{graphicx} %
\usepackage{epsfig,amsmath}
\usepackage{amssymb}

\usepackage{subfig}
\usepackage{rotate}
\usepackage{color}
\usepackage{bm}

\DeclareFontFamily{OT1}{pzc}{}
\DeclareFontShape{OT1}{pzc}{m}{it}%
            {<-> s * [1.10] pzcmi7t}{}
\DeclareMathAlphabet{\mathscr}{OT1}{pzc}%
                                {m}{it}

\definecolor{RedWine}{rgb}{0.743,0,0}
\definecolor{RoyalBlue}{rgb}{0.25,.41,.88}

\def\bm#1{\textbf{\em #1}}

\def\khat{\hat{\bm{k}}}
\def\kk{\bm{k}}
\def\qq{\bm{q}}
\def\bfPsi{\mathbf{\Psi}}
\def\bmx{\bm{x}}

\newcommand{\be}{\begin{equation}}
\newcommand{\ee}{\end{equation}}
\newcommand{\bea}{\begin{eqnarray}}
\newcommand{\eea}{\end{eqnarray}}
\def\ba#1\ea{\begin{align}#1\end{align}}

\newcommand{\refeq}[1]{Eq.~(\ref{eq:#1})}          
\newcommand{\refeqs}[2]{Eqs.~(\ref{eq:#1})--(\ref{eq:#2})}          
\newcommand{\reffig}[1]{Fig.~\ref{fig:#1}}          
\newcommand{\refsec}[1]{Sec.~\ref{sec:#1}}

\newcommand{\vs}{\nonumber\\}

\title[Toward Accurate Modeling of the Nonlinear Bispectrum]{Toward Accurate Modeling of  {the Nonlinear} 
Matter Bispectrum:
 {Standard Perturbation Theory and Transients from Initial Conditions}
}
\author[N. McCullagh et al.]{
Nuala McCullagh,$^{1}$
Donghui Jeong,$^{2,3}$
and Alexander S. Szalay$^{4}$
\\
$^{1}$Institute for Computational Cosmology, Department of Physics, Durham University, South Road, Durham DH1 3LE, UK\\
$^{2}$Department of Astronomy and Astrophysics, The Pennsylvania State University, University Park, PA 16802, USA\\
$^{3}$Institute for Gravitation and the Cosmos, The Pennsylvania State University, University Park, PA 16802, USA\\
$^{4}$Henry A. Rowland Department of Physics and Astronomy, The Johns Hopkins University 3400 N Charles St., Baltimore, MD 21218, USA
}

\begin{document}
\label{firstpage}
\pagerange{\pageref{firstpage}--\pageref{lastpage}}
\maketitle


\begin{abstract}
  Accurate modeling of nonlinearities in the galaxy bispectrum, the 
 Fourier transform of the galaxy three-point correlation function, 
 is essential to fully exploit it as a cosmological probe.
 In this paper, we present numerical and theoretical
 challenges in modeling the nonlinear bispectrum.
 First, we test the robustness of the matter bispectrum measured from 
 $N$-body simulations using different initial conditions generators. We run a suite of $N$-body simulations using the Zel'dovich approximation and second-order Lagrangian 
 perturbation theory (2LPT) at different starting redshifts, and find that 
 transients from initial decaying modes systematically reduce the 
 nonlinearities in the matter bispectrum. To achieve $1\%$ accuracy in 
 the matter bispectrum at $z\le3$ on scales $k< 1$ $h$/Mpc, 
 2LPT initial conditions generator with initial redshift 
 $z\gtrsim 100$ is required.
 We then compare various analytical formulas and empirical fitting 
 functions for modeling the nonlinear matter bispectrum, and discuss the 
 regimes for which each is valid.
 We find that the next-to-leading order (one-loop) correction from standard 
 perturbation theory matches with $N$-body results on quasi-linear scales for $z\ge 1$. We find that the fitting formula
 in \citet{hgm2012} accurately predicts the matter bispectrum for $z\le1$ on a wide range of scales, but at higher redshifts, the fitting formula given in \citet{sc2001} gives the best agreement with measurements from $N$-body simulations.
\end{abstract}

\begin{keywords}
cosmology: theory -- large-scale structure of Universe
\end{keywords}

\section{Introduction}

The large-scale structure of matter and galaxies in the Universe 
is an excellent cosmological probe. Thus far, the two-point correlation
function of galaxies has been widely employed to study the large-scale 
structure, and  broaden our understanding of the Universe by providing 
tight constraints on, e.g., the curvature of the Universe, initial 
conditions, and properties of dark energy (e.g. \citep{sdss2006, wigglez2011, wigglez2012, boss2012, vipers2013}).
A complete description of non-Gaussian large-scale structure, 
however, must include higher-order statistics. Although the matter density
fluctuations follow \textit{nearly} Gaussian statistics at early times, the 
nonlinear process of gravitational instability generically drives the matter 
density non-Gaussian. In addition, a nonlinear relation (bias) between the 
galaxy density and matter-density contrast and the distortion of the 
measured redshift of galaxies due to peculiar velocity further enhance the 
non-Gaussianity.
The leading order statistic sensitive to such non-Gaussianity is
the three-point correlation function, or its Fourier transform the bispectrum.

The galaxy three-point function and bispectrum have been measured previously 
in IRAS \citep{scoccimarroIRAS}, 2dFGRS \citep{verde2dfgrs, gazta2dF2005}, 
SDSS \citep{nicholSDSS2006, nishimichi2007, marinSDSS2011, mcbrideSDSS2011a, mcbrideSDSS2011b, bispectrumDR11}, and WiggleZ \citep{marin2013}, but have 
not been as fruitful as their two-point counterparts because theoretical modeling of higher-point statistics has not 
yet reached the level that is required for sophisticated data analysis.
Future surveys promise orders of magnitude improvement in both galaxy
number density and volume coverage, permitting better determination of the 
higher-order statistics of the cosmic density field, and 
the huge resources that the community has put into those large surveys compels
us to exploit the higher order statistics. 
Upon accessing all the cosmological information, the three-point statistics 
will be a powerful probe of inflation 
\citep{jeongkomatsu2009, sefusatti2007, baldauf2011b},
nonlinear structure formation \citep{scoccimarro1998, scoccimarro1999}, and 
astrophysics such as galaxy formation \citep{baldauf2011b}.

In order to extract information from the bispectrum of galaxies we must have an accurate model of the bispectrum of the underlying matter field.
Cosmological $N$-body simulations are the most direct way of 
studying the nonlinear evolution of the matter bispectrum.
One consideration when modeling the bispectrum using simulations
(in particular at high redshift) is transients from initial conditions 
\citep{scoccimarro1998b, crocce2006}. 
The initial positions and velocities of the $N$-body particles are often 
generated by using Lagrangian Perturbation Theory (LPT), either to first 
order (Zel'dovich approximation) or second order (2LPT).
Because the initial conditions from the Zel'dovich approximation and 2LPT 
include spurious decaying modes---which do not exist in the real 
Universe---on top of the desired fastest growing modes, 
one must wait a long enough time for 
the decaying modes to be sufficiently suppressed \citep{scoccimarro1998b}.
For the case of the power spectrum, where the Zel'dovich transients appear at non-linear order, 2LPT initial conditions with starting
redshift of $z_{\rm ini}=49$ may be enough for capturing the nonlinearities 
at $z=0$ \citep{crocce2006}. 
As we shall show below, however, correct simulation of the matter bispectrum 
at redshifts $1<z<6$ requires even earlier starting redshifts with 2LPT 
initial conditions.
This is because in the Zel'dovich approximation, the decaying modes in the matter bispectrum affect the leading 
order, and suppress the resulting matter bispectrum even on the very 
largest scales in the simulations.

Parallel to numerical simulations is theoretical study of the 
nonlinear evolution of the matter density field. Standard perturbation 
theory (PT) (for a review: \citet{bernardeau2002}) models nonlinearities 
in the matter density field as a pressureless, single fluid evolving under 
gravitational interaction. In PT, the leading order, tree-level, matter 
bispectrum can be calculated from the second order solution for the density 
field, giving us an expression that is valid on large scales.
Beyond tree-level, one can include successive higher order
corrections to the bispectrum, which may correctly model the non-linear
bispectrum in the quasi-linear regime. Other approaches, such as resummed Lagrangian Perturbation Theory (LPT) have also been developed to model the bispectrum on quasi-linear scales \citep{rampf2012}.
For even smaller scales, where the r.m.s. of the density contrast 
is of order unity, PT breaks down and the matter density field is in
the fully nonlinear regime. 
Various phenomenological 
fitting formulas based on $N$-body simulations have been developed that claim to better model the nonlinear behavior of the bispectrum at low redshifts. There are two versions of 
fitting formulas available in the literature:
the first proposed by \citet{sc2001} and more recently one proposed by 
\citet{hgm2012}.
In light of the recent development of renormalized perturbation
theory \citep{croccescoccimarro2006}, one can also try simplistic models 
of `renormalizing' the power spectrum while keeping the tree-level vertex 
(the second order kernel) intact. That is, we assume that the nonlinear 
behavior of the bispectrum is completely described by the nonlinear power 
spectrum, so the usual tree-level expression is altered by replacing the 
linear power spectrum with the nonlinear power spectrum.
To verify these formulas, we compare the three nonlinear models 
of the matter bispectrum with the results from $N$-body simulations with 
the smallest transient effect, and find the region of validity for each of 
model, in particular at high redshifts ($1<z<6$).

In this paper, we shall address the aforementioned 
numerical and theoretical challenges on accurately modeling the matter 
bispectrum in the following order.
In \refsec{visualization}, we discuss various ways 
of visualizing the bispectrum, and introduce a flattening plot,
which shows all of the measured bispectrum on a range of scales, that we shall use throughout 
this paper. 
\refsec{theory} reviews the standard perturbation theory approach to 
calculating the tree-level bispectrum and leading-order transients from initial 
conditions. 
\refsec{transient} presents the effect of transients on the measured bispectrum
from a suite of $N$-body simulations with different initial redshifts and 
initial condition generators (Zeldovich and 2LPT).
We then discuss in \refsec{NLmodeling} theoretical and empirical models of 
the nonlinear bispectrum. We show that the fitting formulae in 
\cite{sc2001,hgm2012} accurately predict the bispectrum in different regimes: \citet{sc2001} is accurate on the widest range of scales and redshifts, and works particularly well at high redshift ($z>1$), whereas \citet{hgm2012} is the most accurate model at lower redshifts ($z\le1$), but over-predicts the bispectrum at high redshifts. We also compare these models to perturbation theory predictions for the bispectrum.
We conclude in \refsec{conclusion}.

\section{Visualization}\label{sec:visualization}
The bispectrum 
 {$B(\kk_1,\kk_2,\kk_3)$ of the Fourier space density contrast 
field $\delta(\kk)$} is defined as:
\begin{align}
\langle \delta(\kk_1)\delta(\kk_2)\delta(\kk_3)\rangle&\equiv (2\pi)^3
B(k_1, k_2,k_3)\delta_D(\kk_1+\kk_2+\kk_3)\text{ ,}\label{eq:bispectrumdef}
\end{align}
where the $\delta_D$ is a Dirac-delta function. The bispectrum is 
non-zero only where the three wavevectors form a triangle due to 
statistical homogeneity of the matter density field. 
Furthermore, statistical isotropy dictates that the bispectrum is independent
of the orientation of the triangle, so the magnitudes of the wavevectors
specify the bispectrum\footnote{With redshift-space distortions, 
statistical istotropy is violated, and the bispectrum depends on the angle 
between the wavevectors and the line-of-sight direction. In this paper,
we only discuss the matter density field in real space.}.
Without loss of generality, we impose the condition that 
$k_1\geq k_2 \geq k_3$ throughout the paper.
For later use, it is useful to consider various limiting triangular 
configurations for reference:
squeezed ($k_1 \approx k_2 \gg k_3$), 
elongated ($k_1 = k_2 + k_3$), 
folded ($k_1 = 2k_2 = 2k_3$), 
isosceles ($k_2 = k_3$), 
and equilateral ($k_1 = k_2 = k_3$).
For graphical illustration of each triangular configuration, we refer
the readers to Fig. 1 of \citep{jeongkomatsu2009}.

\begin{figure}
\includegraphics[width=0.49\textwidth]{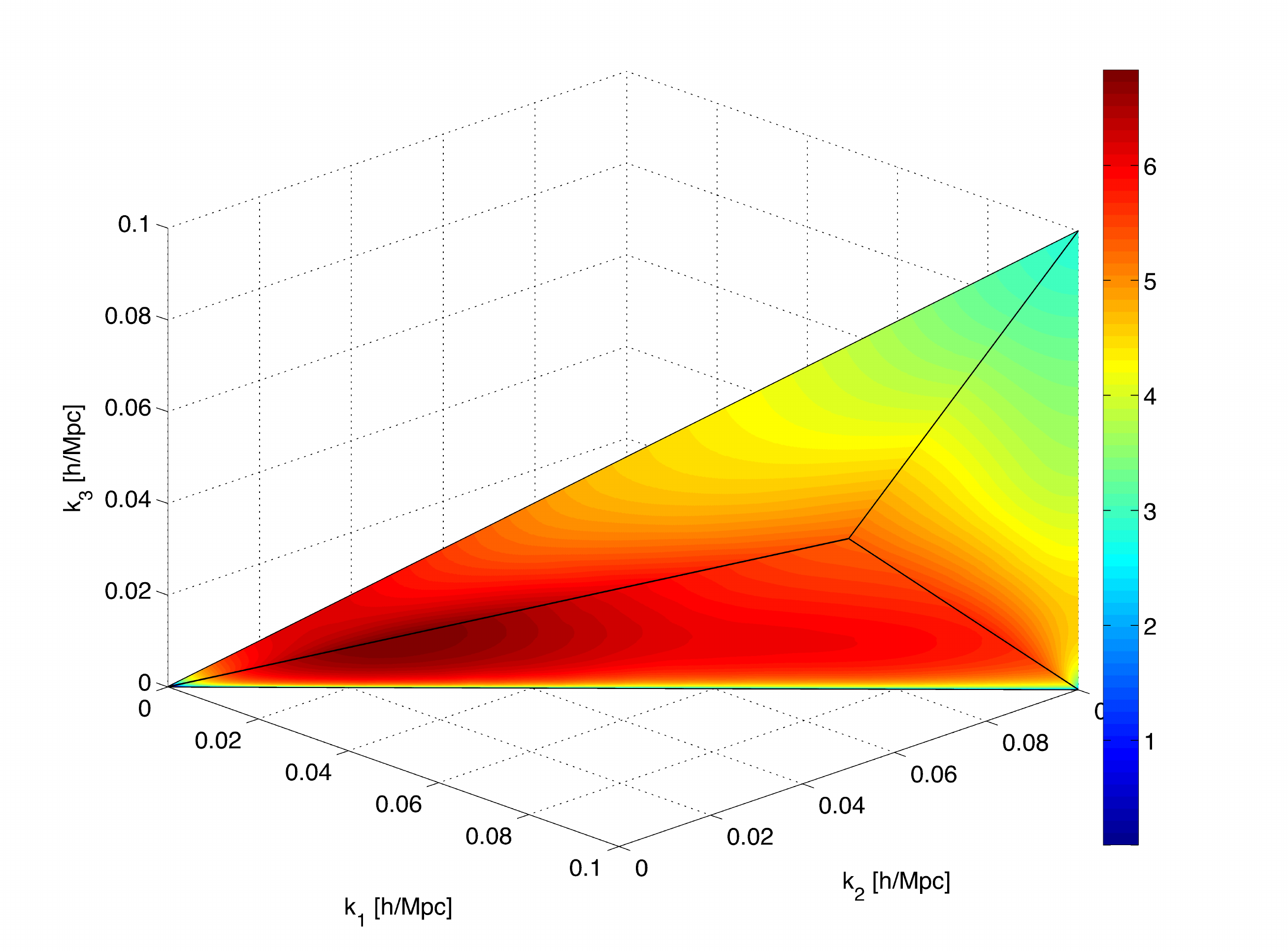}
\caption{
3-D plot of the full tree-level matter bispectrum, with $k_1\geq k_2\geq k_3$.The color reflects the magnitude of $\log\left(B(k_1, k_2, k_3)\right)$.
}
\label{fig:3d1}
\end{figure}

How should we visualize the bispectrum?
\reffig{3d1} shows the leading order matter bispectrum
as a three-dimensional color plot.
As we shall show in the next section, 
the leading order matter bispectrum
in PT is given by
\be
B(k_1,k_2,k_3)
=
2F_2^{(s)}(\kk_1,\kk_2)P_L(k_1)P_L(k_2) + (2~{\rm cyclic})
\label{eq:Bktree}
\ee
where
\be
F_2^{(s)}(\kk_1,\kk_2)
=
\frac57 + \frac{\kk_1\cdot\kk_2}2\left(\frac{1}{k_1^2}+\frac{1}{k_2^2}\right)
+
\frac27 \frac{(\kk_1\cdot\kk_2)^2}{k_1^2k_2^2},
\ee
is the second order PT kernel and $P_L(k)$ is the linear matter power spectrum.
The three axes of \reffig{3d1} represent the magnitudes of the three wave numbers, 
$k_1$, $k_2$, and $k_3$ in which the bispectrum is shown only for the 
tetrahedronic region defined by $k_1 \geq k_2 \geq k_3$ and
$k_1\le k_2+k_3$ (trianglular condition).
The amplitude of the bispectrum is coded by colors.
We can see from Fig \ref{fig:3d1} that the bispectrum has the largest 
magnitude at around $k_1\simeq k_2\simeq k_3\sim 0.02~h/\mathrm{Mpc}$
where the linear power spectrum peaks.
We can also examine the behavior of the bispectrum at the limiting triangular 
configurations, which are along the planes and lines that make up the shape 
of the region in which the bispectrum is defined.
This way of visualization is often used in the cosmic microwave background (CMB)
literature when discussing the bispectrum of temperature anisotropies
\citep{fergusson2012, planckbispectrum}.
While the three-dimensional color plots are useful for showing the overall 
structure of the bispectrum, it is not apt for showing its detailed shape and scale
dependence. 

\begin{figure*}
\begin{center}
\begin{subfloat}[$k_1=0.01$ h/Mpc]{
\includegraphics[scale=0.4]{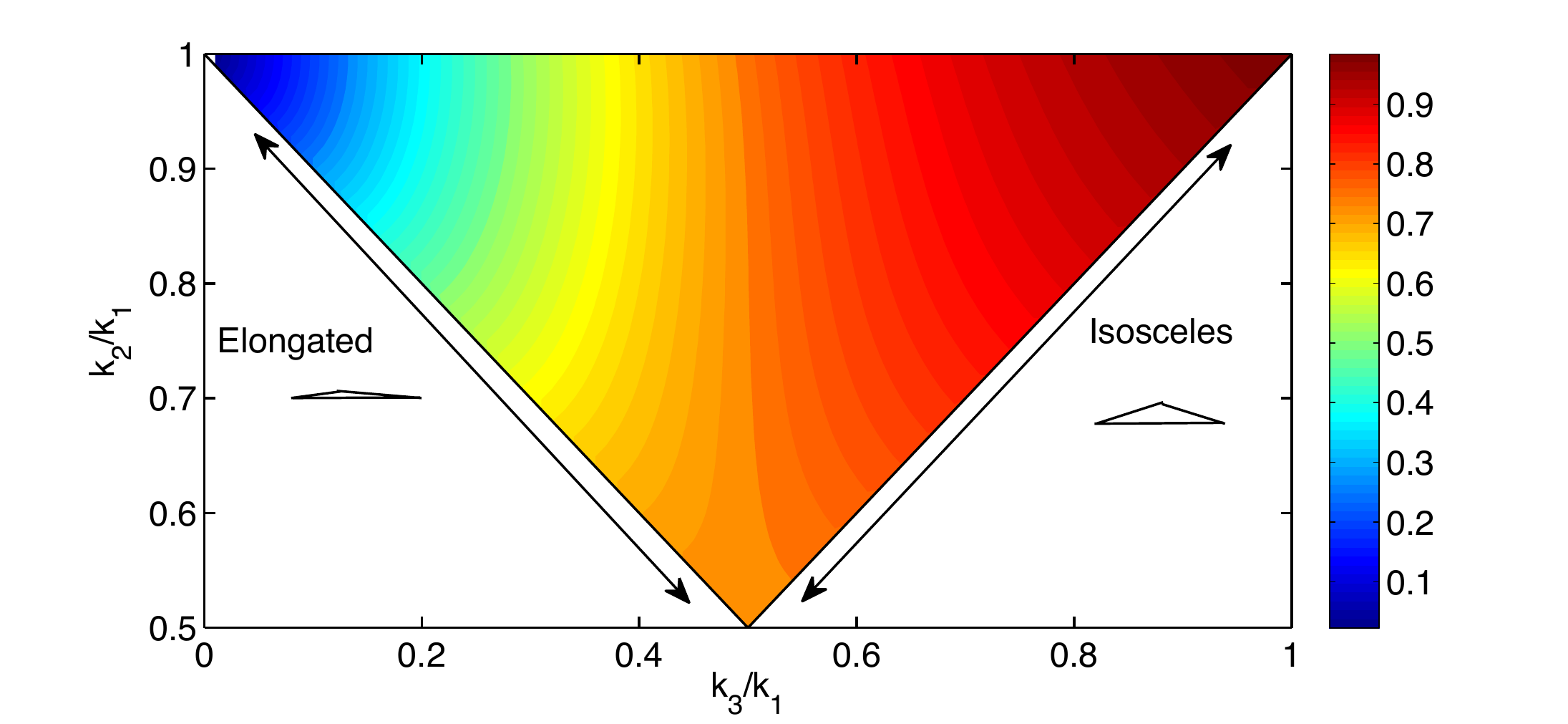}}
\end{subfloat}
\begin{subfloat}[Flattened bispectrum, with $\delta k_1=0.001$ h/Mpc]{
\includegraphics[scale=0.35]{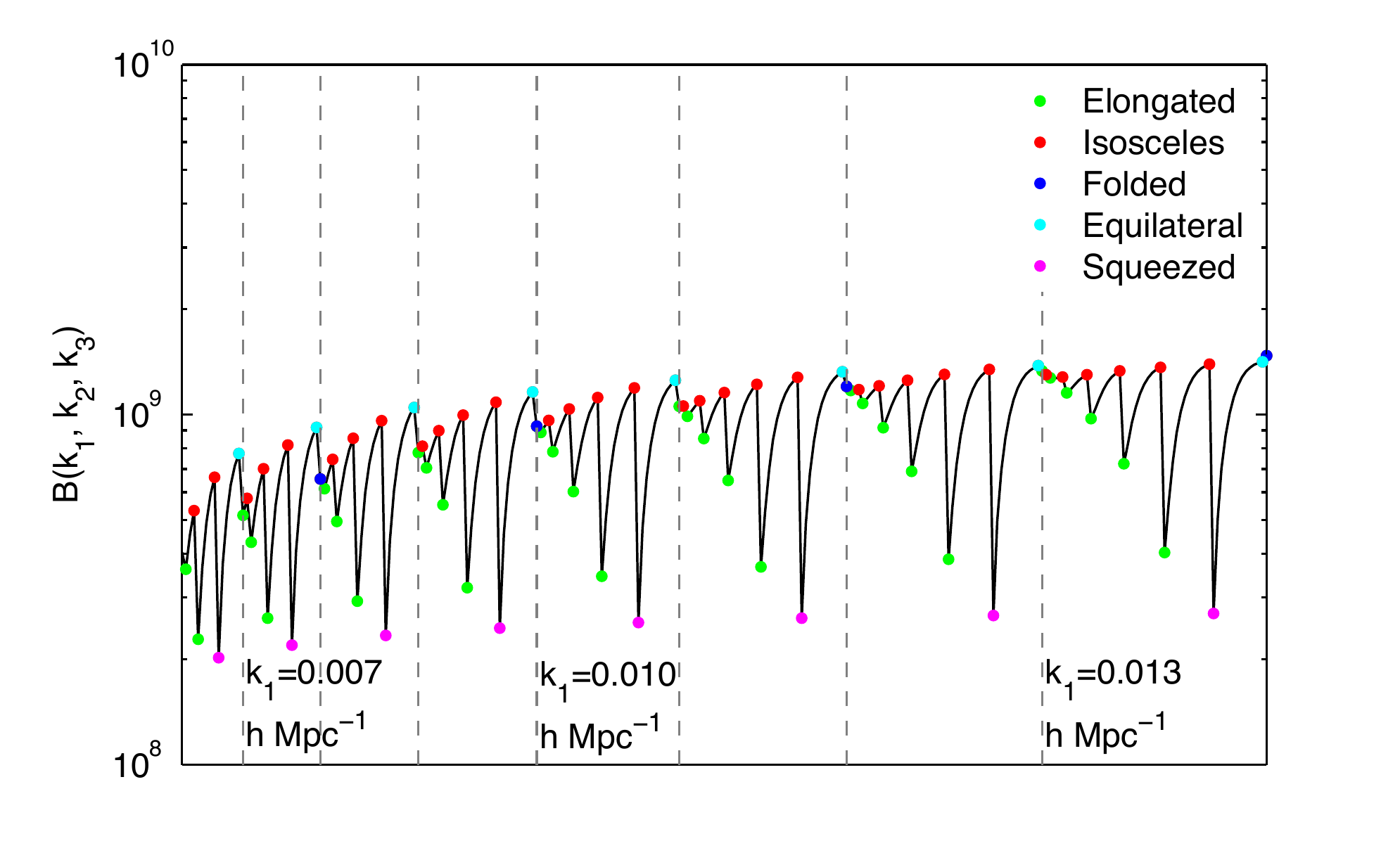}}
\end{subfloat}
\\
\begin{subfloat}[$k_1=0.05$ h/Mpc]{
\includegraphics[scale=0.4]{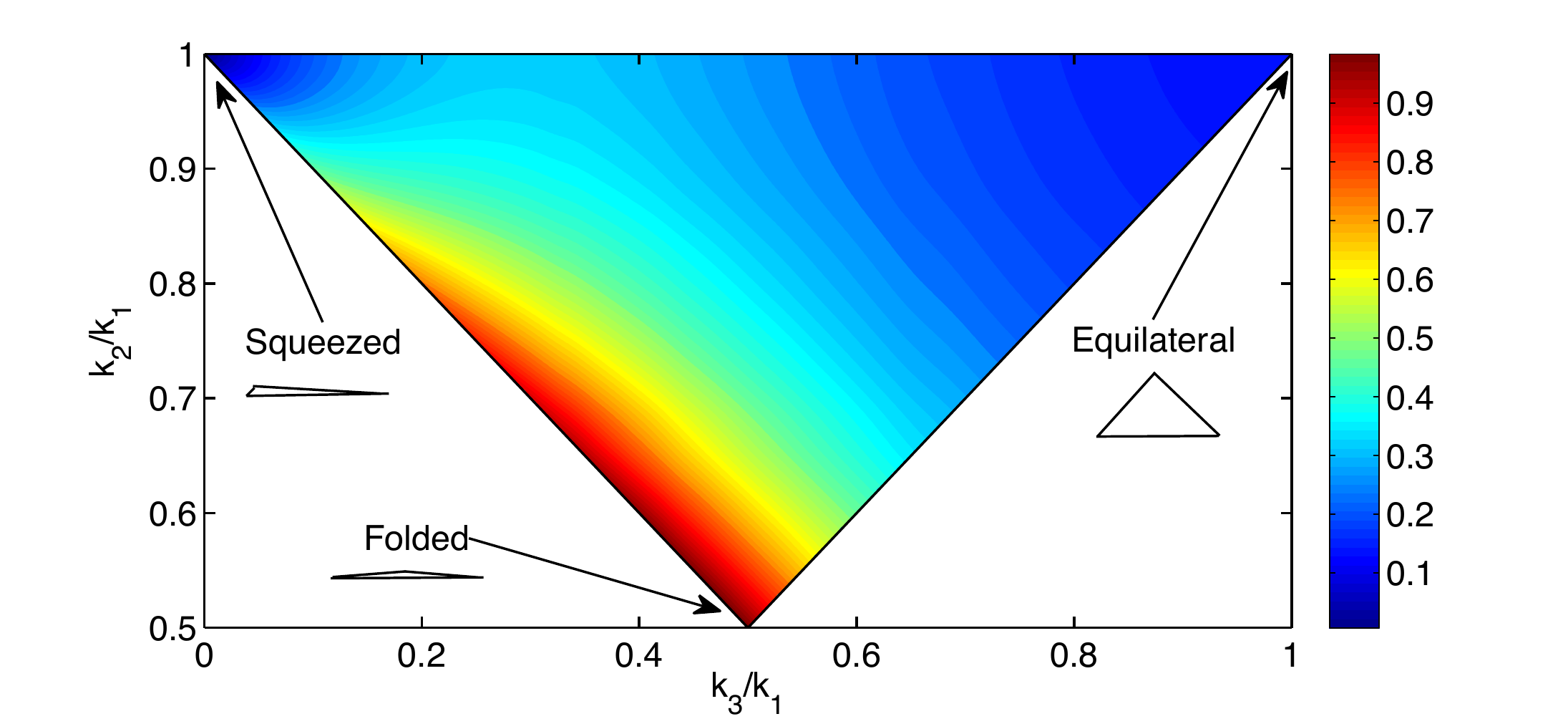}}
\end{subfloat}
\begin{subfloat}[Flattened bispectrum, with $\delta k_1=0.001$ h/Mpc]{
\includegraphics[scale=0.35]{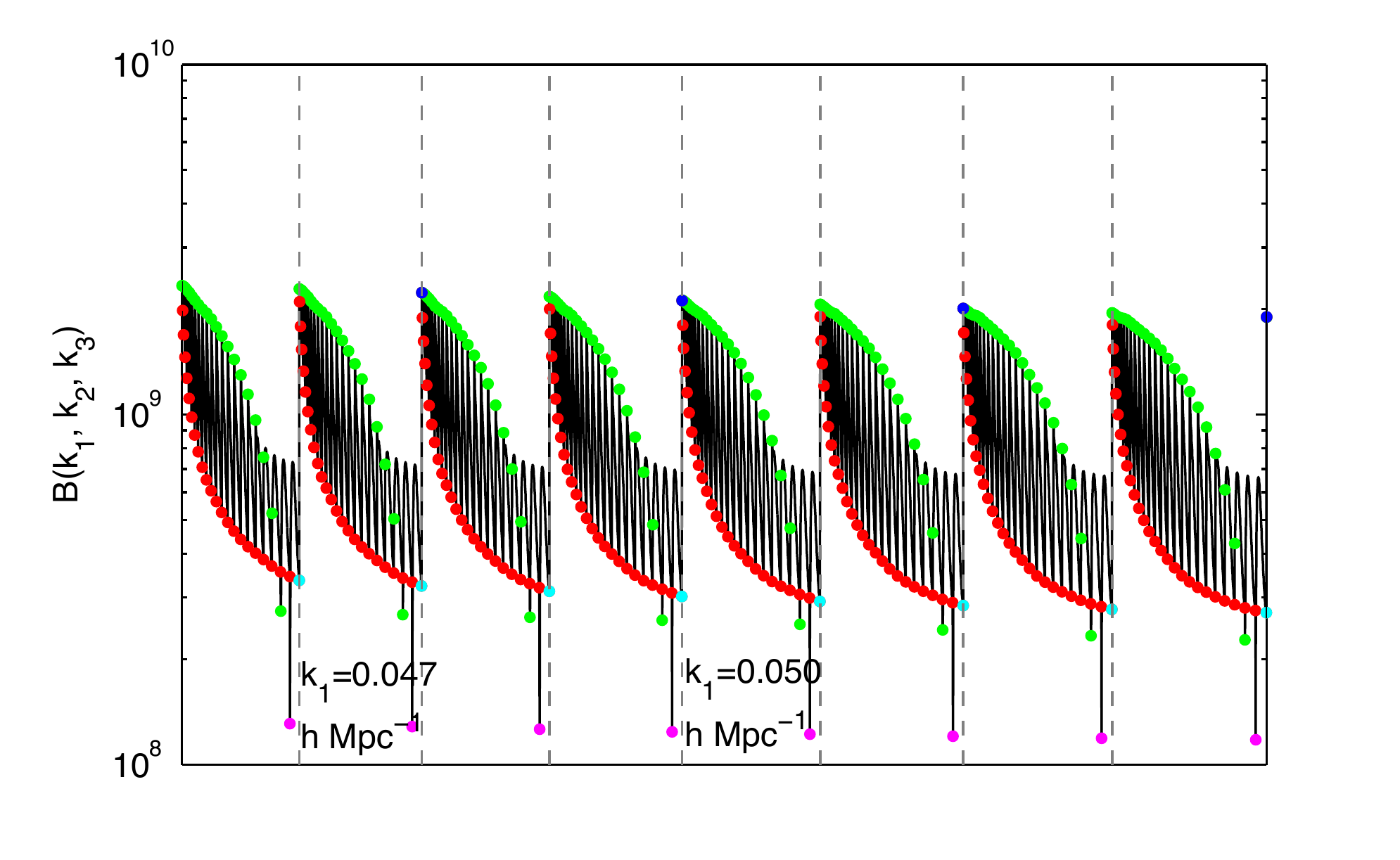}}
\end{subfloat}
\\
\begin{subfloat}[$k_1=0.15$ h/Mpc]{
\includegraphics[scale=0.4]{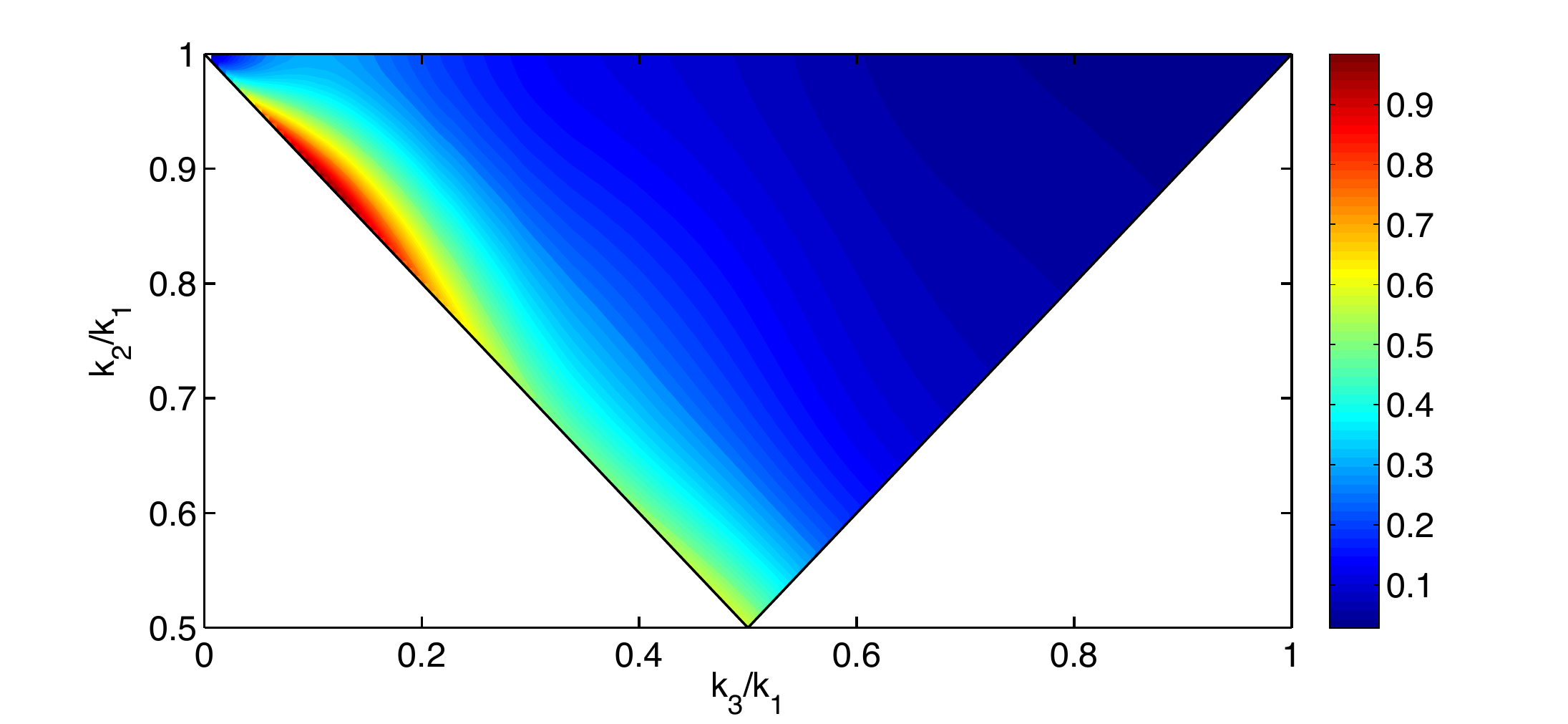}}
\end{subfloat}
\begin{subfloat}[flattened bispectrum, with $\delta k_1=0.005$ h/Mpc]{
\includegraphics[scale=0.35]{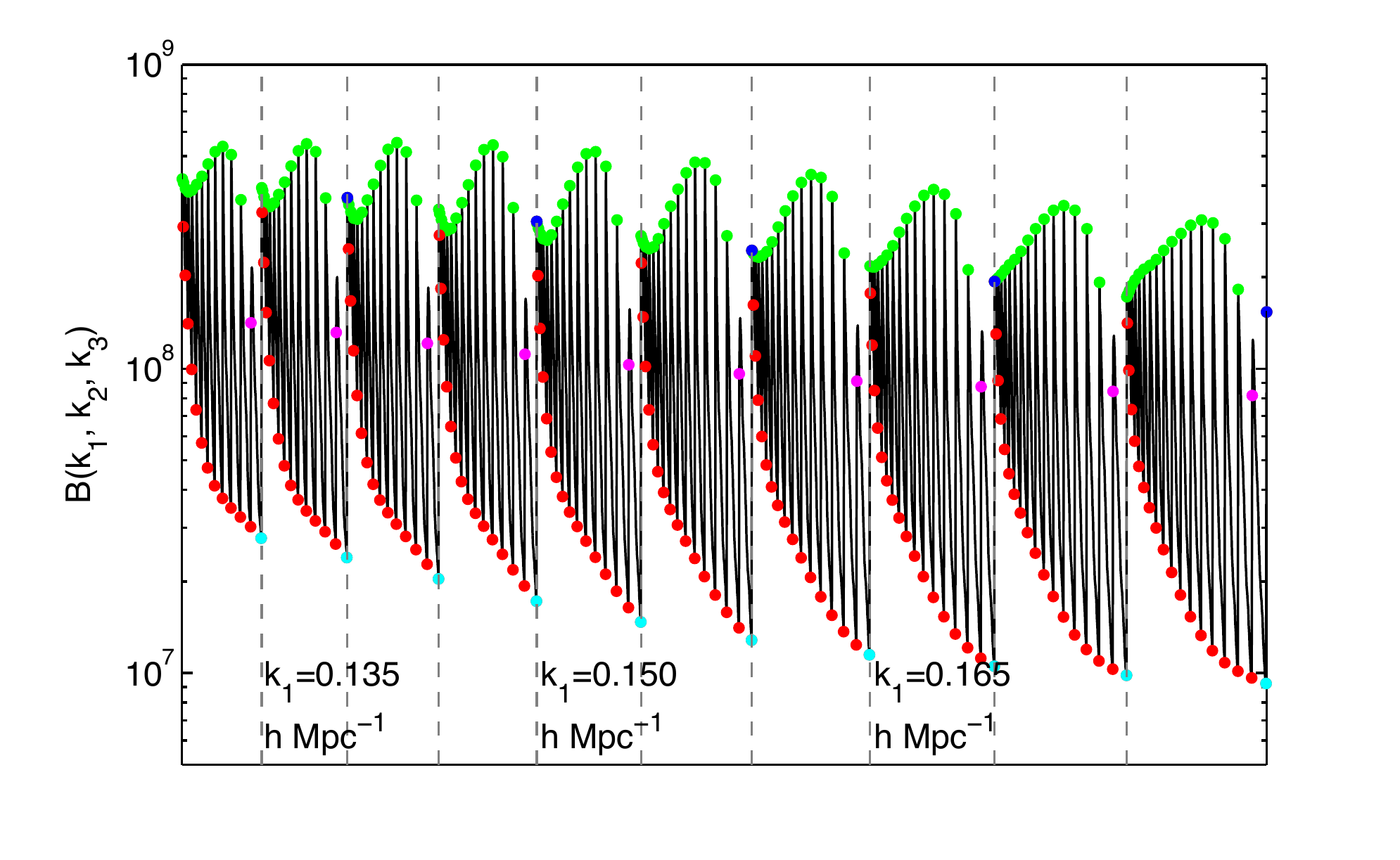}}
\end{subfloat}
\caption{Left: slices of the bispectrum at constant $k_1$. The locations of different triangular configurations are highlighted. Right: the flattened bispectrum with a given $\delta k_1$, where the range of $k_1$ plotted includes the corresponding slice on the left. Vertical dashed lines indicate different slices of $k_1$. Values of $k_1$ in several slices is given in units of $h$/Mpc. }
\label{fig:k2k3color2}
\end{center}
\end{figure*}

A useful way of visualizing the shape and scale dependence of 
bispectrum is by plotting slices of constant $k_1$. 
The plots in the left column of Fig \ref{fig:k2k3color2} show slices of the bispectrum  {in \reffig{3d1}}
at $k_1=0.01$ $h$/Mpc, $k_1=0.05$ $h$/Mpc, and $k_1=0.15$ $h$/Mpc
(from top to bottom). Note that the values shown have been normalized by the maximum value of $B(k_1, k_2, k_3)$ in each slice.
We also show the locations of the different triangular configurations in
the same plot. 
From these plots, we can see that the squeezed configuration always has a 
suppressed signal compared to other elongated triangles. On large scales, 
the bispectrum of equilateral triangles is higher than in other 
configurations, whereas on smaller scales, it is suppressed 
compared to, say, isosceles triangles. \citet{jeongkomatsu2009}
present a detailed discussion about the shape and scale dependence of the 
leading order matter bispectrum.

In reality, we can only measure the Fourier space density 
contrast at discrete grid points whose separation is determined by the 
volume of the $N$-body simulation, or the width of the survey window function.
The bispectrum then can also be measured only for a finite number of triples 
$(k_1,k_2,k_3)$. We can facilitate the comparison among different predictions 
for the bispectrum by flattening all possible triplets
into a one-dimensional vector ($x$-axis), where we rank the triplet
in row-major, ascending order with the condition $k_1\geq k_2\geq k_3$.
This allows us to look at all of the triangular configurations over a range 
of scales simultaneously, as well as compare bispectra on the $y$-axis. 
Each plot in the right column of Fig. \ref{fig:k2k3color2} shows the 
bispectrum of all triangular configurations in a range of $k_1$ that 
includes the slice shown in the left panel. 
The $x$-axis shows the ranked $(k_1,k_2,k_3)$ triplets, and the $y$-axis 
shows the amplitude of bispectrum. As one can see from the plots, a similar 
pattern block is duplicated for each value of the long-wavelength mode $k_1$, 
as indicated by vertical dashed lines.
We also show the configuration dependence by highlighting the limiting 
triangles in different colors. One can easily relate this plot to the 
contour plot in the left panel. For a given $k_1$, the first-ranked triplet 
is at the bottom vertex ($k_2=k_3=k_1/2$, blue dots, folded triangle)
of the triangular region in the left panels ($k_3/k_1$-$k_2/k_1$ plane).
The rank is followed by the next elongated triangle (green dots) 
then moves rightward to the isosceles triangles (red dots).
This process continues until the rank reaches most-squeezed triplet
then walks along the upper-most horizontal line to the equilateral 
triplet. Depending on the scales ($k_1$ values) that we are looking at, 
the configuration dependence of the bispectrum is different.

In our analysis for this paper, we shall mainly use this flattened 
representation to study the configuration and scale dependence of 
bispectrum at all triangles. Although non-linearities change the exact shape 
of the bispectrum `chunk' for a given $k_1$, the overall structure remains 
unchanged.

\section{Transients in Standard Perturbation Theory}\label{sec:theory}
In this section, we review the standard perturbation theory (PT)
formalism for calculating the tree-level bispectrum \citep{bernardeau2002}.
We then follow the approach of \citet{scoccimarro1998b} and 
\citet{crocce2006} to 
compute the correct growing and decaying modes for initial conditions set 
using the Zel'dovich Approximation and 2LPT initial conditions.

One way of simplifying the single fluid equations is combining
the density field $\delta(\kk,\eta)$ and velocity divergence field 
$\theta(\kk,\eta)$ into a doublet as:
\be
\Psi_a(\kk,\eta)=\left(\delta(\kk, \eta),-\theta(\kk,\eta)/\mathcal{H}\right)\ 
\ee
where the logarithm of the linear growth factor ($D(\tau)$)
relative to its initial value ($D_i$): $\eta \equiv \ln (D(\tau)/D_i)\ $ 
is the time variable.
Here, $\mathcal{H}\equiv d\ln a/d\tau =  aH$ is the reduced Hubble parameter
that is defined with conformal time $\tau=\int dt/a(t)$.
The equations of motion for $\Psi_a$, combining the continuity equation, 
the Euler equation, and the Poisson equation, can then be written as\footnote{
As the initial conditions are set in the deeply matter-dominated epoch, and 
transient effects are most dominant at higher redshifts, our analysis
throughout this paper is restricted to a flat, matter-dominated 
(Einstein de-Sitter) universe. Later time dark energy will only marginally
change the analysis as the PT kernels are insensitive to the background 
cosmology \cite{bernardeau2002}.}:
\begin{align}
\partial_{\eta}\Psi_a(\kk,\eta)&+\Omega_{ab}\Psi_b(\kk,\eta)\notag\\
&=\gamma^{(s)}_{abc}(\kk,\kk_1,\kk_2)\Psi_b(\kk_1,\eta)\Psi_c(\kk_2,\eta)\label{eq:motion}.
\end{align}
Here, 
\be
\Omega_{ab}\equiv\begin{bmatrix} \ 0 & -1\  \\-3/2 & 1/2\  \end{bmatrix}
\ee
is the linear mixing matrix, and 
\begin{align}
\gamma_{121}^{(s)}(\kk,\kk_1,\kk_2)&=(2\pi)^3\delta_D(\kk-\kk_1-\kk_2)\frac{\kk\cdot \kk_1}{2 k_1^2}\\
\gamma_{222}^{(s)}(\kk,\kk_1,\kk_2)&=(2\pi)^3\delta_D(\kk-\kk_1-\kk_2)\frac{k^2(\kk_1\cdot\kk_2)}{2 k_1^2k_2^2}
\end{align}
encode the non-linear interactions. 
Note that $\gamma_{121}^{(s)}(\kk,\kk_1,\kk_2)=\gamma_{112}^{(s)}(\kk,\kk_2,\kk_1)$ and 
the integration $\int d^3k_i/(2\pi)^3$ over the repeated wavevectors
($\kk_1$ and $\kk_2$) are implied.

A Laplace transform of Equation (\ref{eq:motion}) in the variable $\eta$ 
leads to:
\begin{align}
&\omega\Psi_a(\kk, \omega)-\phi_a(\kk)+\Omega_{ab}\Psi_b(\kk,\omega)\label{eq:laplace1}\\
=&\gamma^{(s)}_{abc}(\kk,\kk_1,\kk_2)\int_{c-i\infty}^{c+i\infty} \frac{d \omega_1}{2\pi i}\Psi_b(\kk_1,\omega_1)\Psi_c(\kk_2,\omega-\omega_1)\ ,\notag
\end{align}
where
\begin{align}
\phi_a(\kk)=\Psi_a(\kk, \eta=0)
\end{align}
is the density and velocity field at the initial time (i.e. the time at which 
we start the simulation). The linear part of \refeq{laplace1} can be 
written in terms of a matrix $\sigma_{ab}$:
\begin{align}
\sigma_{ab}^{-1}(\omega)&\equiv \omega\delta_{ab}+\Omega_{ab}\\
\sigma_{ab}(\omega)&=\frac{1}{(2\omega+3)(\omega-1)}\begin{bmatrix} 2\omega+1 & 2\   \\ \ 3 & 2\omega\  \end{bmatrix},
\end{align}
and, finally, we obtain the formal solution through 
an inverse Laplace transformation as:
\begin{align}
&\Psi_a(\kk,\eta)=g_{ab}(\eta)\phi_b(\kk)\label{eq:solution}\\
&+\int_0^{\eta} d\eta'g_{ab}(\eta-\eta') \gamma_{bcd}^{(s)}(\kk,\kk_1,\kk_2)\Psi_c(\kk_1,\eta')\Psi_d(\kk_2,\eta') \notag.
\end{align}
Here,
\ba
g_{ab}(\eta)&=\int_{c-i\infty}^{c+i\infty}\frac{d\omega}{2\pi i} \sigma_{ab}(\omega)e^{\omega \eta}\label{eq:propagator}\\
&=\frac{e^{\eta}}{5}\begin{bmatrix} 3 & 2\   \\ \ 3 & 2\  \end{bmatrix}-\frac{e^{-3\eta/2}}{5}\begin{bmatrix} -2 & 2\   \\ \ 3 & -3\  \end{bmatrix}\notag
\ea
is the Green's function which describes the time evolution of the linear 
perturbations. The first term proportional to $e^\eta\propto a$ is the 
growing mode and the second term proportional to $e^{-3\eta/2}\propto a^{-3/2}$ 
is the decaying mode.

\refeq{solution} is the \textit{formal} solution of the fluid equations, and 
starting points of various of \textit{renormalized} perturbation theories. 
For our purposes, however, it suffices to find the perturbative solution 
that follows. We perturbatively expand the solutions at initial 
and final time ($\phi_a$ and $\Psi_a$, respectively) in terms of the linear 
(Gaussian) density $\delta_0(\kk)$ field at initial time as:
\begin{align}
\phi_a(\kk)&=\sum_{n=1}^{\infty}\phi_a^{(n)}(\kk)\\
\Psi_a(\kk,\eta)&=\sum_{n=1}^{\infty}\Psi_a^{(n)}(\kk,\eta)
\end{align}
where
\begin{align}
\phi_a^{(n)}(\kk)&=[\delta_D]_n\mathcal{T}_a^{(n)}(\kk_1,\cdots,\kk_n)\delta_0(\kk_1)\cdots\delta_0(\kk_n) \label{eq:pert} \\
\Psi_a^{(n)}(\kk,\eta)&=[\delta_D]_n\mathcal{F}_a^{(n)}(\kk_1,\cdots,\kk_n, \eta)\delta_0(\kk_1)\cdots\delta_0(\kk_n) 
\end{align}
and $[\delta_D]_n\equiv \delta_D(\kk-\kk_1-\cdots-\kk_n)$. 
Here, $\mathcal{T}_a^{(n)}$ are the $n$-th order kernels of the initial 
density and velocity fields, and $\mathcal{F}_a^{(n)}(\eta)$ are the 
$n$-th order resulting kernels at a later time $\eta$.
Again, the integration $\int d^3k_i/(2\pi)^3$ over the repeated $\kk_i$'s 
is implied.
By using this ansatz, we calculate the kernels $\mathcal{F}_a^{(n)}$ at the 
final redshift from the initial kernels $\mathcal{T}_a^{(n)}$:
\begin{align}
\label{eq:finalkernel_recursion}
\mathcal{F}_a^{(n)}(\eta)&=g_{ab}(\eta)\mathcal{T}_b^{(n)}\\
&+\sum_{m=1}^{n-1}\int_0^{\eta}ds \ g_{ab}(\eta-s)\gamma_{bcd}^{(s)}\mathcal{F}_c^{(m)}(s)\mathcal{F}_d^{(n-m)}(s)\notag.
\end{align}
This recursion relation is the starting point of the transient analysis.

\subsection{Growing mode solutions of Standard Perturbation Theory}\label{sec:PTsolution}
In the real universe, the decaying modes decay away at very early times, 
and the density field is at its fastest growing mode for any reasonable 
choice of the initial redshift.
Therefore, the initial conditions for ideal $N$-body simulations must be 
set at an early enough time such that PT is valid on the smallest resolved 
scales, and set by the fastest growing modes at all relevant orders in PT.

In this section, we find the fastest growing mode solution perturbatively. 
We first find the linear growing mode $\phi_a^{(1)}(\kk)=\delta_0(\kk)(1,1)$ 
so that the decaying part of the Green's function \refeq{propagator} vanishes.
Then, the second order kernels at time $\eta$ are given by 
\refeq{finalkernel_recursion} using the linear growing kernel of 
$\mathcal{F}_a^{(1)} = e^{\eta}(1,1)$.
While the time integral in \refeq{finalkernel_recursion} gives rise to the
slowly growing mode ($\propto e^{\eta}$) and the decaying mode
($\propto e^{-\frac32\eta}$), those are exactly canceled by the initial
fastest-growing kernels $\mathcal{T}_a^{(2)}=(F_2^{(s)},G_2^{(s)})$ defined
by the kernels in standard PT \citep{bernardeau2002} and we are left with the 
fastest-growing solution of 
$\mathcal{F}_2^{(n)}(\eta) = e^{2\eta}(F_2^{(s)},G_2^{(s)})$.
The same is true for all the higher order kernels, and 
the fastest growing solutions are  
\begin{align}
\Psi_a^{(n)}(\kk, \eta)=&(2\pi)^3 \delta_D(\kk-\kk_1-\cdots-\kk_n)
\vs
&\times\mathcal{F}_a^{(n)}(\kk_1,..,\kk_n, \eta)\delta_0(\kk_1)\cdots\delta_0(\kk_n),
\label{eq:idealsolution}
\end{align}
where
\begin{align}
\mathcal{F}_a^{(n)}(\eta) &= e^{n\eta}
\left(
F_n^{(s)},G_n^{(s)}
\right)
\label{eq:idealkernels}
\end{align}
are the standard PT kernels.
In particular, the second order kernel which appears in the leading-order 
matter bispectrum in \refeq{Bktree} is given by
\be
F_2^{(s)}(\kk_1,\kk_2)
=
\frac57 + \frac{\kk_1\cdot\kk_2}2\left(\frac{1}{k_1^2}+\frac{1}{k_2^2}\right)
+
\frac27 \frac{(\kk_1\cdot\kk_2)^2}{k_1^2k_2^2},
\label{eq:F2s}
\ee
and the velocity kernel is given by
\be
G_2^{(s)}(\kk_1,\kk_2)
=
\frac37 + \frac{\kk_1\cdot\kk_2}2\left(\frac{1}{k_1^2}+\frac{1}{k_2^2}\right)
+
\frac47 \frac{(\kk_1\cdot\kk_2)^2}{k_1^2k_2^2}.
\label{eq:G2s}
\ee

When the single fluid approximation is valid 
(at early times and on large scales),
ideal $N$-body simulations must reproduce the density and velocity fields 
in \refeq{idealsolution} with the growing mode kernels \refeq{idealkernels}. 
To satisfy this condition, we must start simulations with initial density 
and velocity fields satisfying \refeq{idealsolution}.

\subsection{Transients from Initial conditions of $N$-body simulations}
Generating the fastest-growing initial conditions at all orders in 
perturbation theory is quite non-trivial. 
This is in part because we simulate $N$-body dynamics 
with the motion of matter \textit{particles}. 
That is, the initial conditions for $N$-body simulations
are set by position and velocity of each particle (rather than set by density 
and velocity gradient field as is demanded by the PT analysis).
Due to the nonlinear relation between particle's position and 
the resulting density field, the initial conditions can only satisfy 
the fastest-growing condition perturbatively, in an order-by-order manner.
As a result, when generating initial conditions with the growing mode at 
a given order in $\delta_0$, it is inevitable that spurious decaying 
modes will be excited at higher orders.

In this section, we study the decaying modes for two of the most 
widely used initial condition generators: first order 
(Zel'dovich approximation) and second order (2LPT) Lagrangian 
perturbation theory (LPT).
For each initial condition generating scheme, we calculate the initial 
density and velocity kernels $\mathcal{T}_a^{(n)}$, and find the evolved kernel
at the final redshift by using \refeq{finalkernel_recursion} in order to 
study the resulting transients in the matter bispectrum as a function of 
initial and final redshifts as well as triangular configuration and scales.

LPT describes the evolution of the matter density field by displacement $\bfPsi$
between each matter particle's initial Lagrantian position $\qq$ 
and final position $\bm{x}$:
\be
\bm{x}(\qq,\eta) = \qq + \bfPsi(\qq,\eta),
\ee
then the peculiar velocity field is given by
\be
\bm{v}(\qq,\eta) 
= \frac{d\bmx}{d\tau}
\equiv \bfPsi'(\qq,\eta).
\ee
Note that $\bmx$ is the comoving coordinate, and prime denotes the time
derivative with respect to the conformal time $\tau$.
When it comes to generating initial conditions of $N$-body simulations,
$\qq$ refers to the position of particles on a regular grid, and 
$\bm{x}$ refers to particle's initial position.

For a given displacement field $\bfPsi$, we find the Fourier space density 
contrast and velocity gradient in terms of the displacement field as
\ba
\delta(\kk,\eta) 
=&
\int d^3 \bm{x} \,\delta(\bm{x}) e^{-i\kk\cdot\bm{x}}
\vs
=& \sum_{n=1}\frac{1}{n!}\int d^3q e^{-i\kk\cdot\qq}
\left[
-i\kk\cdot\bfPsi(\qq,\eta)
\right]^n,
\label{eq:LPTdelta}
\\
\theta(\kk,\eta) 
=& 
\int d^3q 
J(\qq,\eta)
\nabla_{x}\cdot\bfPsi'(\qq,\eta)
e^{-i\kk\cdot(\qq + \bfPsi(\qq,\eta))},
\label{eq:LPTtheta}
\ea
where we calculate the Jacobian $J(\qq,\eta)$ by using mass conservation 
$J(\qq,\eta)=\left|\partial^3 x/\partial^3 q\right|
=\left(1+\delta(\bm{x})\right)^{-1}
=|\delta_{ij}+\bfPsi_{i,j}|$.
Here, comma stands for the derivative with respect to the Lagrangian
coordinate. As the cosmic matter field on large scales is irrotational, 
we introduce the scalar potential such that 
$\Psi_i = -\phi_{,i}$.
In $n$-th order LPT, the scalar potential $\phi$ is expanded up to 
$n$-th order in the linear density contrast $\delta_0(\kk)$ \citep{bouchetLPT}.

\subsubsection{Zel'dovich approximation}
In the Zel'dovich approximation, the scalar potential $\phi$ is given by
\be
\phi(\kk) = -\frac{1}{k^2} \delta_0(\kk),
\ee
and from \refeqs{LPTdelta}{LPTtheta} we find the linear order initial kernels 
is 
\be
\mathcal{T}_a^{\mathrm{ZA},(1)} = (1,1).
\ee
This means the Zel'dovich approximation indeed gives rise to the growing
mode solution in linear order. On the other hand, the second order initial
kernels are
\ba
\mathcal{T}_a^{\mathrm{ZA},(2)} 
= 
\left(
\frac{(\kk\cdot\kk_1)(\kk\cdot\kk_2)}{2k_1^2k_2^2},~
\frac{k^2(\kk_1\cdot\kk_2)}{2k_1^2k_2^2}
\right),
\ea
which are different from the fastest-growing mode kernels of PT 
in \refeq{idealkernels}. Therefore, the transients in the Zeld'ovich 
approximation start to appear from second order, and affect the 
subsequent evolution of the density and velocity fields at second order as
\ba
&F_2^{\mathrm{ZA},(2)}(\kk_1,\kk_2,\eta)
=
e^{2\eta}F_2^{(s)}(\kk_1,\kk_2,\eta)
\vs
&\qquad+
\left(
\hat{t}_{ij}(\kk_1)\hat{t}_{ij}(\kk_2) - \frac23
\right)
\left(
\frac{3}{10}e^{\eta} - \frac{3}{35}e^{-\frac32\eta}
\right)
\\
&G_2^{\mathrm{ZA},(2)}(\kk_1,\kk_2,\eta)
=
e^{2\eta}G_2^{(s)}(\kk_1,\kk_2,\eta)
\vs
&\qquad+
\left(
\hat{t}_{ij}(\kk_1)\hat{t}_{ij}(\kk_2) - \frac23
\right)
\left(
\frac{3}{10}e^{\eta} + \frac{9}{70}e^{-\frac32\eta}
\right).
\ea
Here, $\hat{t}_{ij}(\kk)$ is the longitudinal operator which relates the 
density field and the tidal field $s_{ij}(\kk)$
\be
s_{ij}(\kk)
=
\left[
\frac{k_ik_j}{k^2} - \frac{\delta_{ij}}{3}
\right]\delta(\kk)
\equiv \hat{t}_{ij}(\kk)\delta(\kk),
\ee
which means that the Zel'dovich approximation contributes an extra tidal field
to the $N$-body simulation at second order.

The extra tidal field then yields a transient in the leading order matter 
bispectrum, and the slowest-decaying transient is given by
\ba
&B^{\mathrm{ZA}}(k_1,k_2,k_3)
-
B(k_1,k_2,k_3) 
\vs
=& 
\frac{3}{5}
\left[\frac{D}{D_i}\right]^{3}
\left(
\mu_{12}^2 - 1
\right)
P_L(k_1)P_L(k_2) 
+ (2~\mathrm{cyclic}),
\ea
with $\mu_{ij} =\khat_i\cdot\khat_j$ being the angular cosine between two
 wave vectors. With \refeq{finalkernel_recursion}, one can also show that,
with Zel'dovich initial conditions,
the slowest-decaying (or, longest-lasting) transients at all higher orders 
($n\ge2$) are suppressed only by $e^{\eta}$ with respect to the fastest-growing
modes.

\subsubsection{2LPT}
In 2LPT, the scalar potential is given by 
\ba
\phi(\kk) = 
\frac{\delta_0(\kk)}{k^2}
+
(2\pi)^3
\delta^{D}(\kk-\kk_{12})
\frac{3(1-\mu_{12}^2)}{14k^2}
\delta_0(\kk_1)
\delta_0(\kk_2),
\ea
with $\kk_{12} = \kk_1+\kk_2$, which reads, from \refeqs{LPTdelta}{LPTtheta},
\ba
\mathcal{T}_a^{\mathrm{2LPT},(1)} =& 
\ (1,1)
\\
\mathcal{T}_a^{\mathrm{2LPT},(2)} =&
\ (F_2^{(s)},G_2^{(s)}).
\ea
Then, as shown in \refsec{PTsolution}, the density and velocity fields in 2LPT
remain as the fastest-growing modes, and an $N$-body simulation with 2LPT 
initial conditions correctly reproduces the leading order bispectrum.

On the other hand, 2LPT initial conditions differ from the fastest-growing
modes at third order. This generates spurious
slowly growing modes and decaying modes. Again, using 
\refeq{finalkernel_recursion}, one can show that, for 2LPT initial conditions,
the slowest-decaying modes are suppressed by two powers in the linear 
growth factor ($e^{-2\eta}$) with respect to the fastest growing mode 
at $n(\ge3)$-th order. Therefore, we expect the following time dependence of 
transient:
\be
B^{\mathrm{2LPT}}(k_1,k_2,k_3) - B(k_1,k_2,k_3)
\propto \left[\frac{D}{D_i}\right]^{2},
\ee
and to contribute on smaller scales than the Zel'dovich case because the 
transient only comes from next-to-leading order and beyond.
Relative to the leading-order, fastest growing mode, the transient decays 
faster than the Zel'dovich case as $(D/D_i)^{-2}$.

\section{Transients in Simulations}\label{sec:transient}

In this section, we study the error induced by transients numerically with 
a suite of simulations with different initial redshifts, using both the 
Zel'dovich approximation (ZA) and 2LPT for setting the initial positions and 
velocities of particles. 

We examine a set of twelve simulations with different initial redshifts, but with the same random seed to set initial conditions. This ensures that within each set, any differences we measure 
between the simulations are due solely to the initial conditions generator. 
The initial conditions in six of the runs were generated using ZA, 
and in the other six we used 2LPT. 
We use the following initial redshifts in each case: 
$z_{\rm init}=400, 300, 200, 100, 50, 25$.
We use the 2LPT simulation started at $z_{\rm init}=400$ 
as our reference simulation to which all other runs are compared.
That is, this simulation defines our fiducial statistics, $P^{\mathrm{fid}}$ and 
$B^{\mathrm{fid}}$.

We run the simulations using the TreePM part of the publicly available 
version of  {\sf Gadget 2} code. 
Each simulation is run with $512^3$ particles in a 200 Mpc/$h$ box and 
particle-mesh Fourier grid $N_{\rm grid}=1024^3$. 
The background universe is $\Lambda$CDM with the following cosmological 
parameters ($\Omega_{\Lambda}=0.73$, $\Omega_m=0.27$, $h=0.7$), and the initial density field is calculated from the primordial curvature power spectrum
with spectral index $n_s=0.96$, and normalized so that the r.m.s. fluctuation 
smoothed by a spherical top-hat filter with radius $8~\rm{Mpc}/h$ is
$\sigma_8=0.8$ at present ($z=0$).
To minimize the effect from accumulated numerical noise,
we have required stringent accuracy of N-body simulation by setting 
parameters of {\sf Gadget 2} following the choice of \cite{crocce2006}:
{
ErrTolIntAccuracy = 0.01, 
MaxRMSDisplacementFac = 0.1, 
MaxSizeTimestep = 0.01, 
ErrTolTheta = 0.2, 
ErrTolForceAcc = 0.002}.
We calculate the cloud-in-cell density from the output particle distribution
on a $1024^3$ grid at $z=6, 5, 4, 3, 2$ and $1$ and compute the 
power spectrum and bispectrum at each redshift.

\subsection{Transients in the matter power spectrum}\label{sec:Pktransient}

\begin{figure*}
\centering
\begin{subfloat}[Transient effect at $z=1$]{
\includegraphics[width=0.47\textwidth]{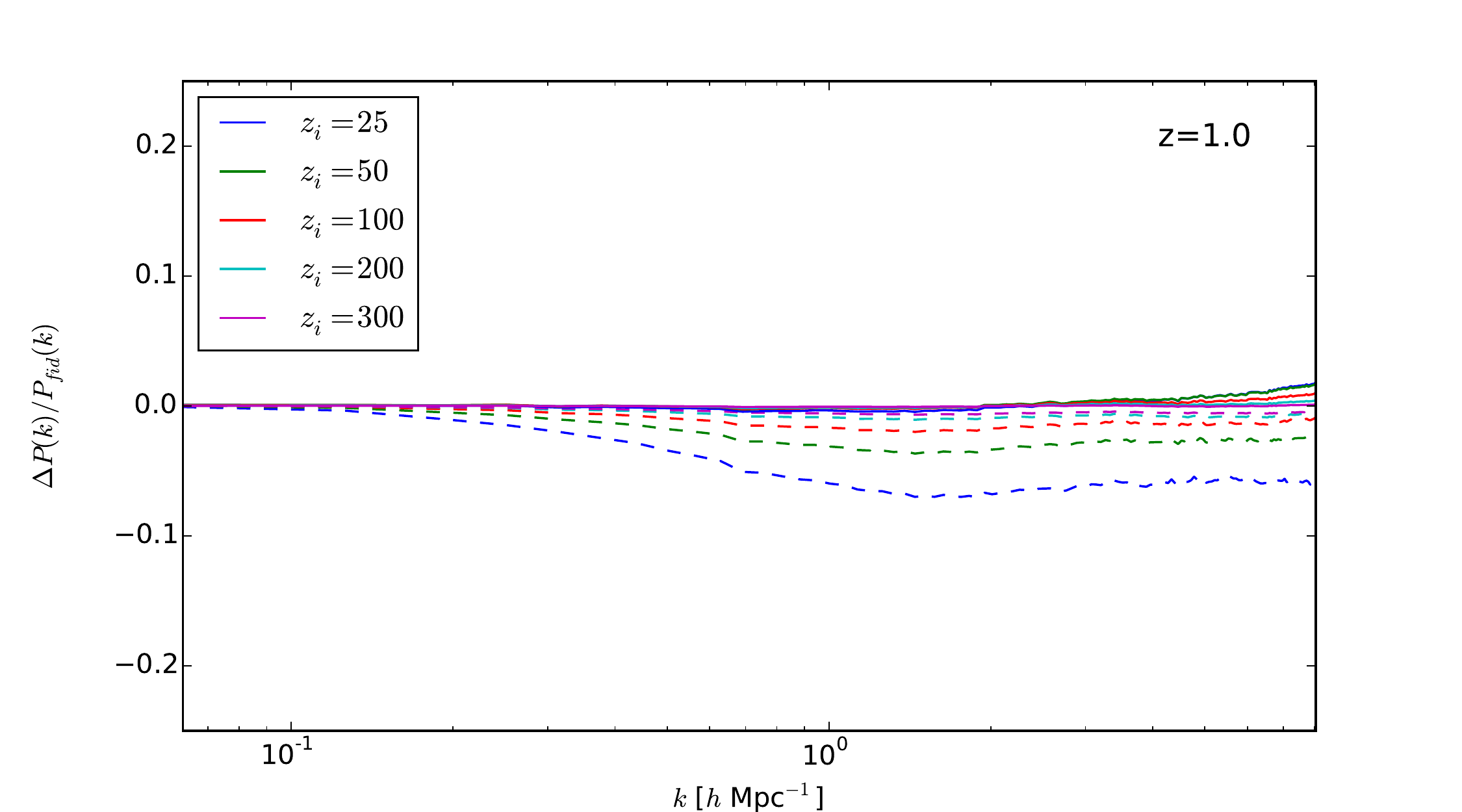}}
\end{subfloat}
\begin{subfloat}[Transient effect at $z=2$]{
\includegraphics[width=0.47\textwidth]{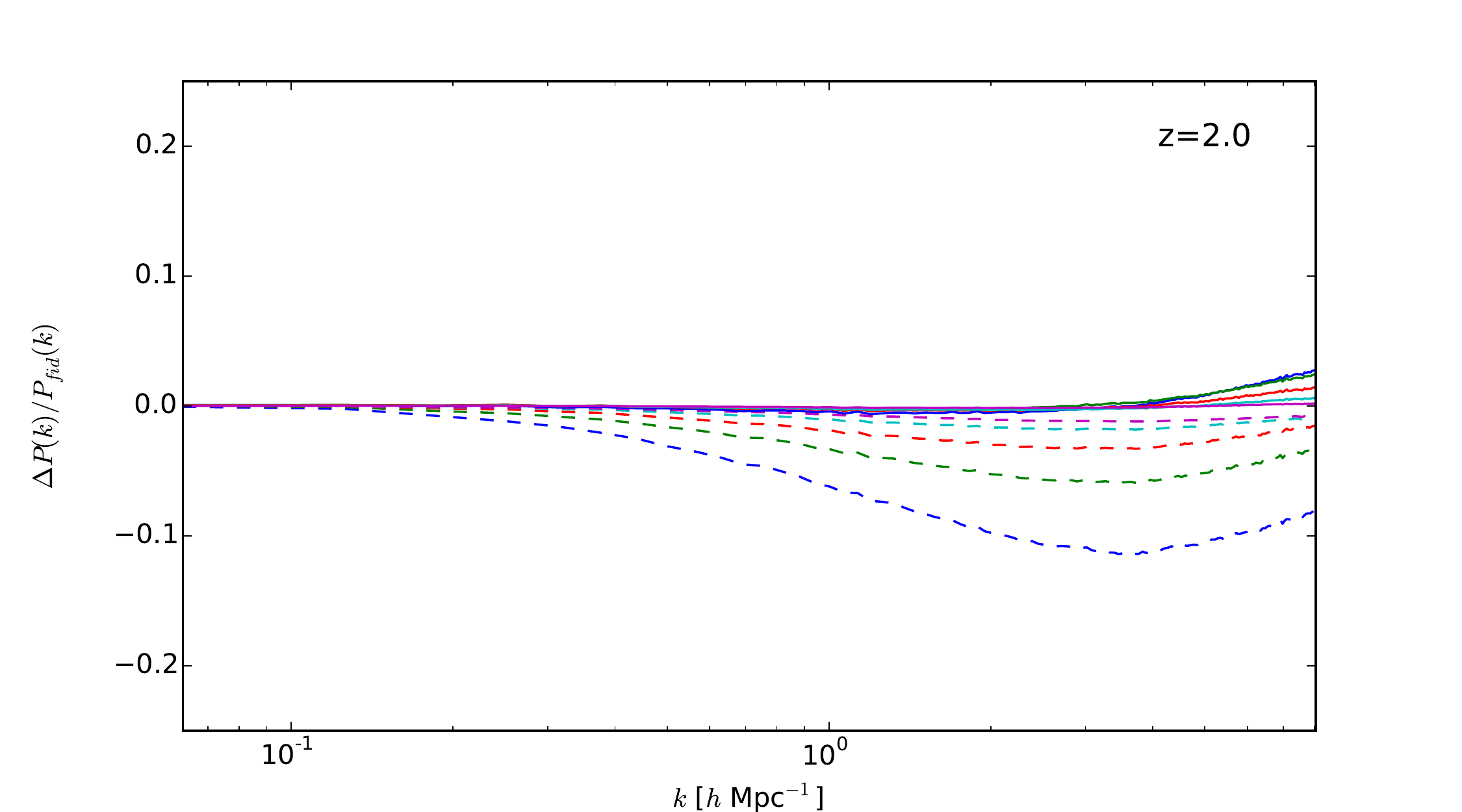}}
\end{subfloat}
\begin{subfloat}[Transient effect at $z=3$]{
\includegraphics[width=0.47\textwidth]{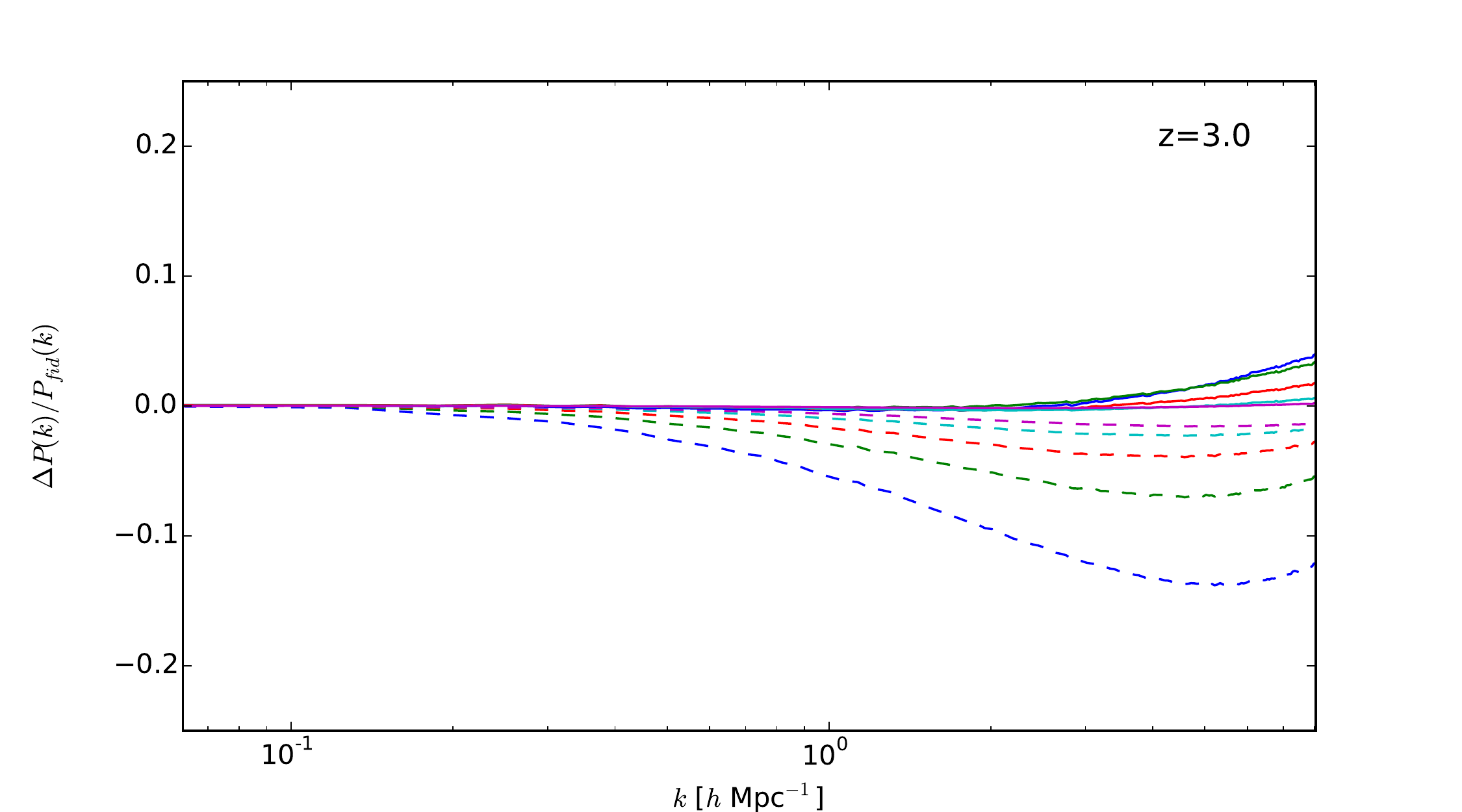}}
\end{subfloat}
\begin{subfloat}[Transient effect at $z=6$]{
\includegraphics[width=0.47\textwidth]{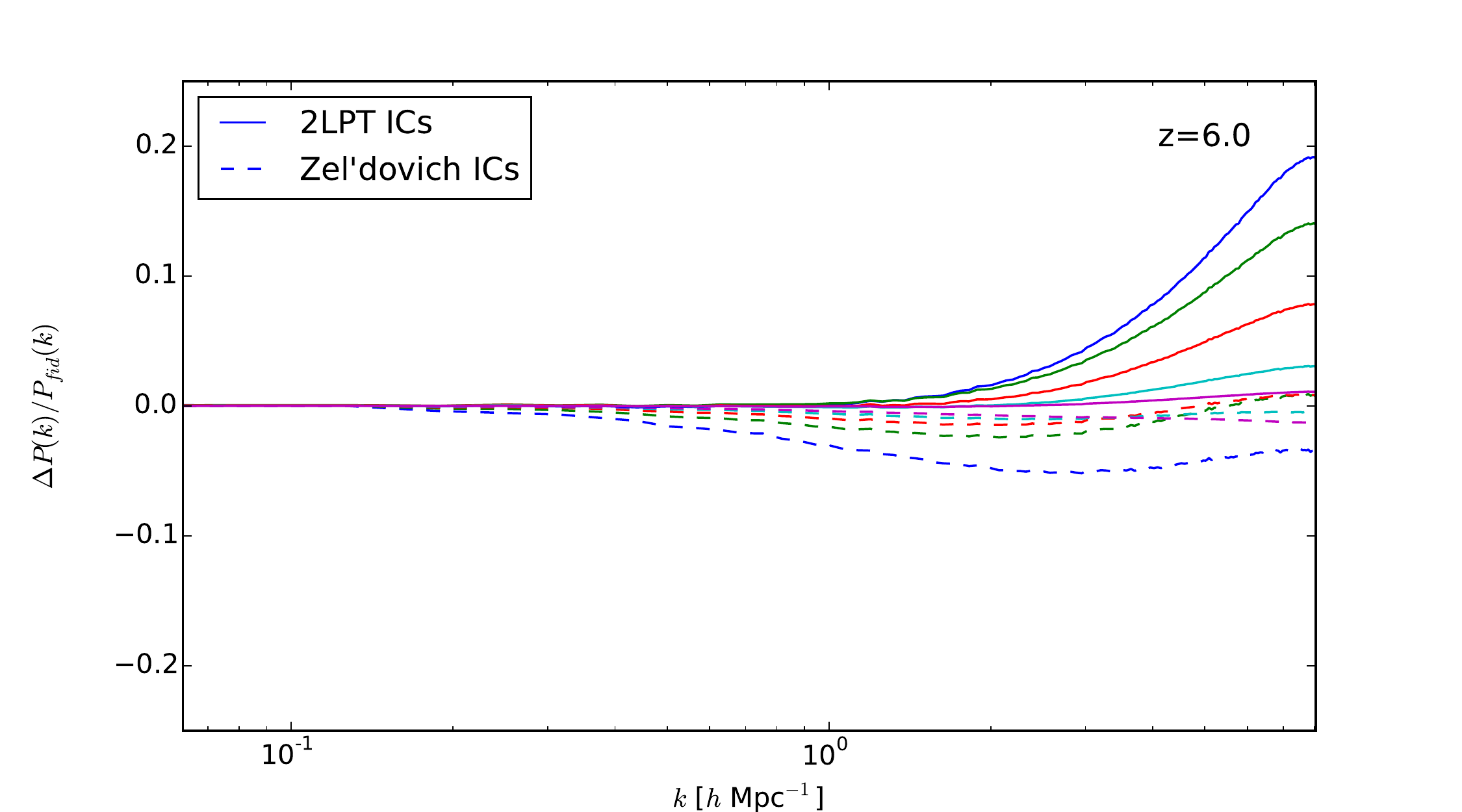}}
\end{subfloat}
\caption{Error induced by transients in the measured power spectrum at redshifts $z=1, 2, 3, 6$ using Zel'dovich and 2LPT initial conditions with different initial redshifts. In each plot, the left column shows the results from the large-box simulation and the right column shows the small-box simulation. The top panels of the plots show the error in simulations using 2LPT initial conditions and the bottom panels show the error from Zel'dovich initial conditions.}
\label{fig:ptransall}
\end{figure*}

Figure \ref{fig:ptransall} shows the fractional error in the measured power spectrum, as compared to the fiducial power spectrum, at $z=1, 2, 3$, and $6$, from 2LPT (solid) and ZA (dashed) initial conditions. The various initial redshifts are shown in different colors. As expected, the transients from ZA initial conditions are in general larger than 2LPT on all scales and redshifts, apart from on very small scales at $z=6$, and the error is larger for smaller initial redshifts in all cases. While the 2LPT power spectrum transients decay quickly, the ZA transients remain significant, even on relatively large scales, until $z=1$. These findings are consistent with PT predictions for  transients in the matter power spectrum, and also with previous studies \cite{crocce2006,jeong:2010}.

\subsection{Transients in the matter bispectrum}\label{sec:Bktransient}
\begin{figure*}
\centering
\begin{subfloat}[Transient effect at $z=1$]{
\includegraphics[width=0.47\textwidth]{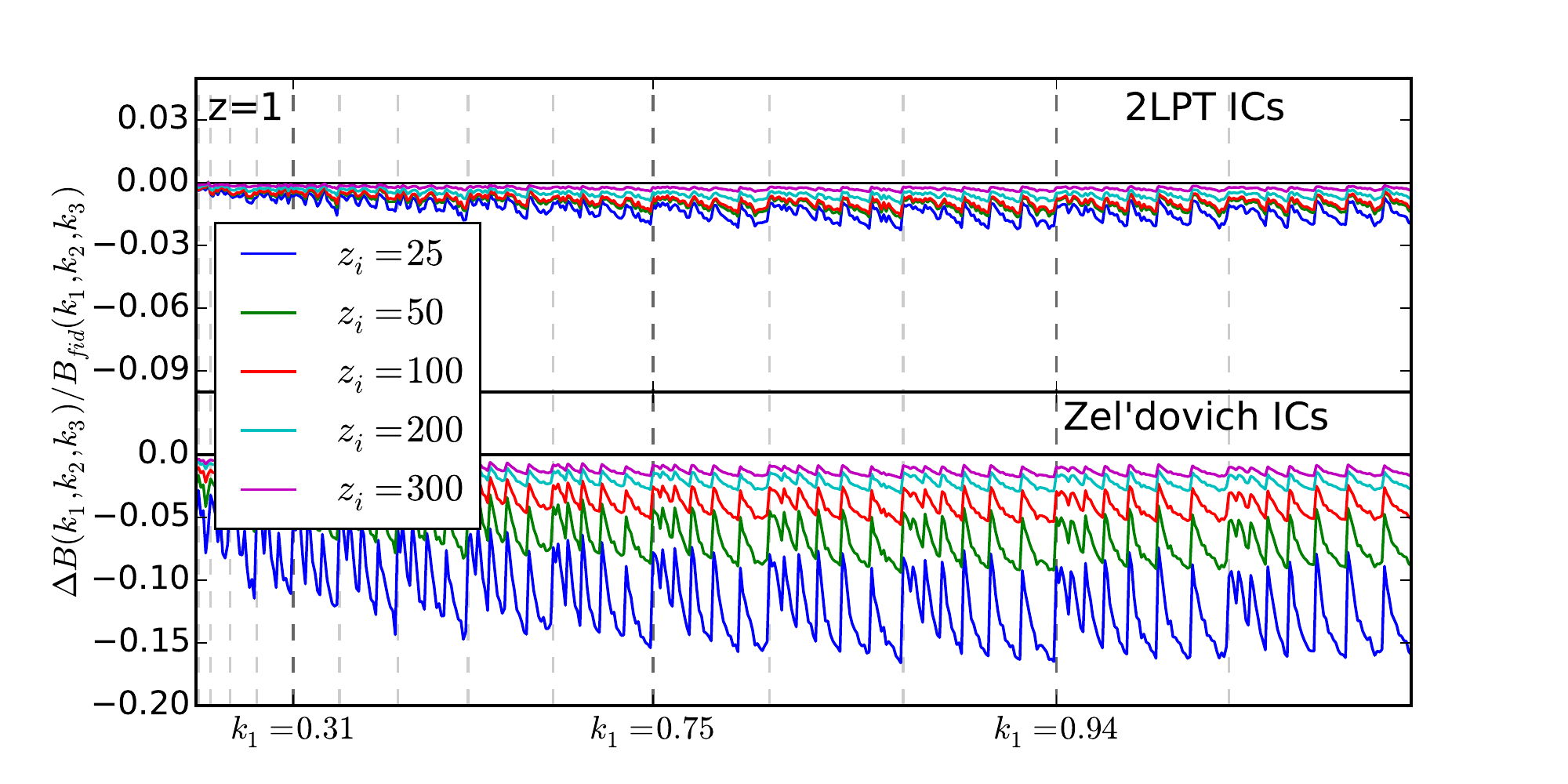}}
\end{subfloat}
\begin{subfloat}[Transient effect at $z=2$]{
\includegraphics[width=0.47\textwidth]{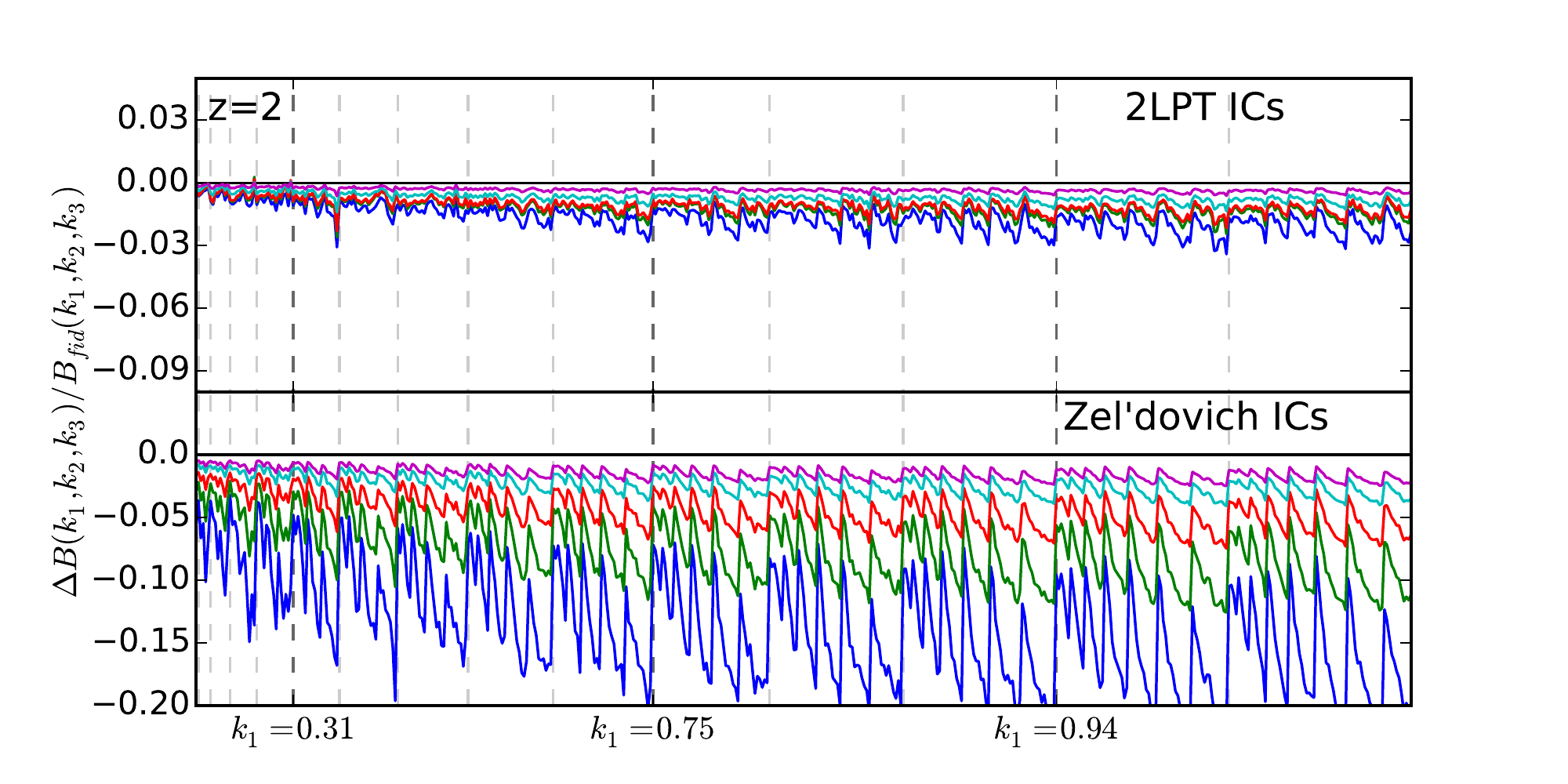}}
\end{subfloat}
\begin{subfloat}[Transient effect at $z=3$]{
\includegraphics[width=0.47\textwidth]{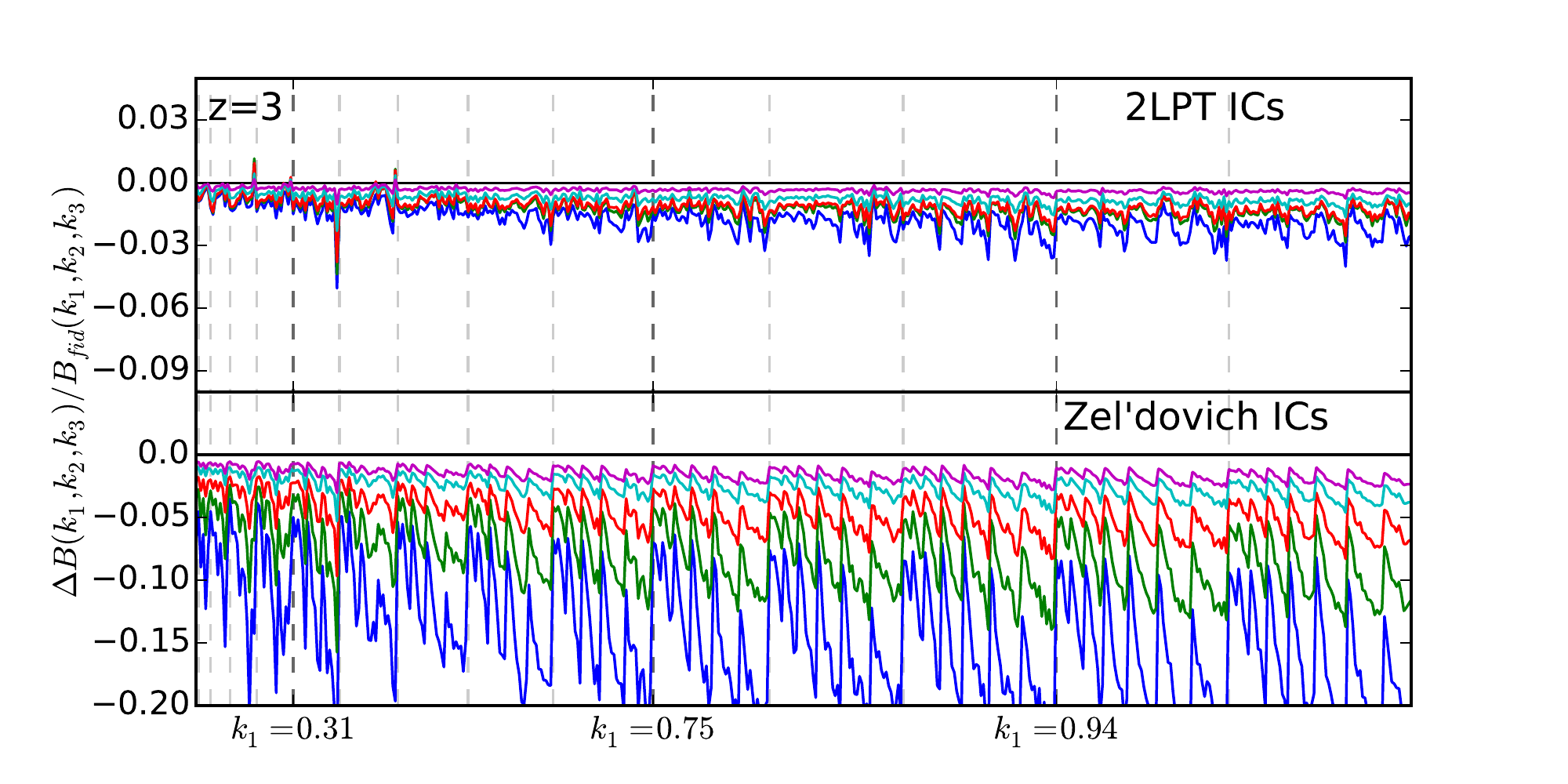}}
\end{subfloat}
\begin{subfloat}[Transient effect at $z=6$]{
\includegraphics[width=0.47\textwidth]{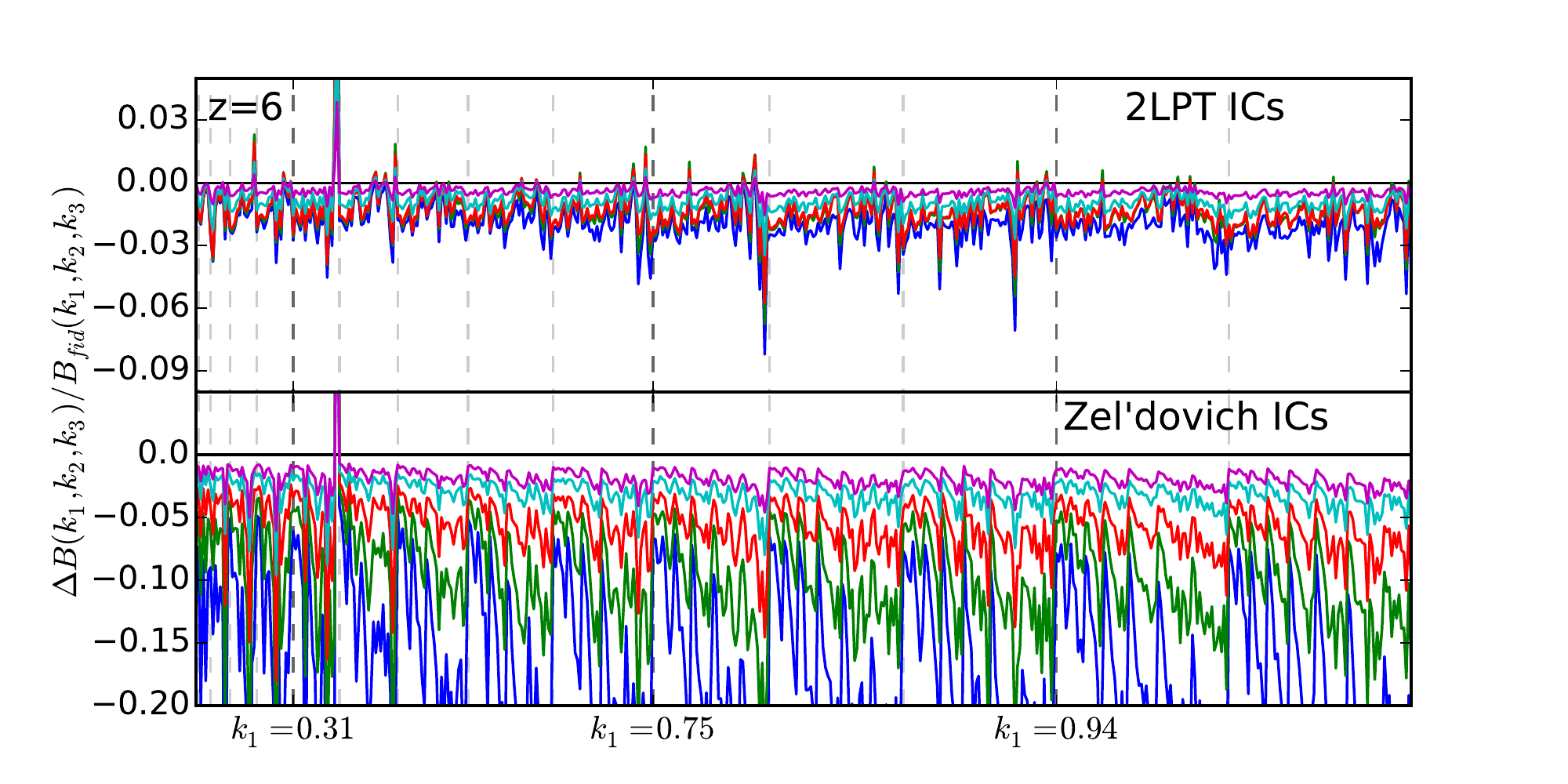}}
\end{subfloat}
\caption{Error induced by transients in the measured bispectrum at redshifts $z=1, 2, 3, 6$ using Zel'dovich and 2LPT initial conditions with different initial redshifts. The top panels of the plots show the error in simulations using 2LPT initial conditions at various initial redshifts and the bottom panels show the error from Zel'dovich initial conditions. The different line colors indicate different initial redshifts for the simulations. Vertical dashed lines indicate slices of $k_1$, and the $k_1$ values of several slices are indicated along the $x$-axis in units of $h$/Mpc.}
\label{fig:btransall}
\end{figure*}

\begin{figure*}
\centering
\begin{subfloat}[Transient effect at $z=1$]{
\includegraphics[width=0.47\textwidth]{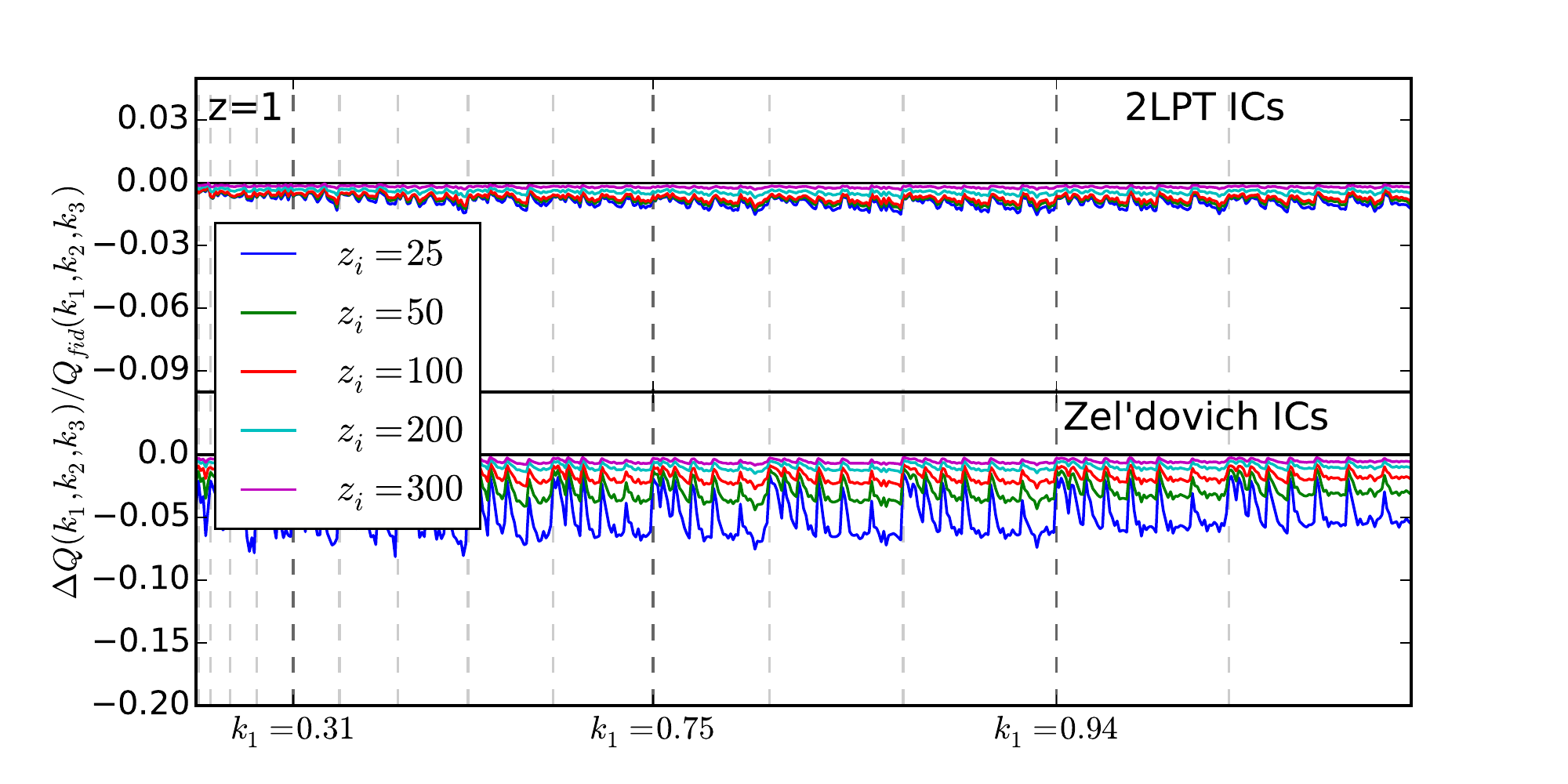}}
\end{subfloat}
\begin{subfloat}[Transient effect at $z=2$]{
\includegraphics[width=0.47\textwidth]{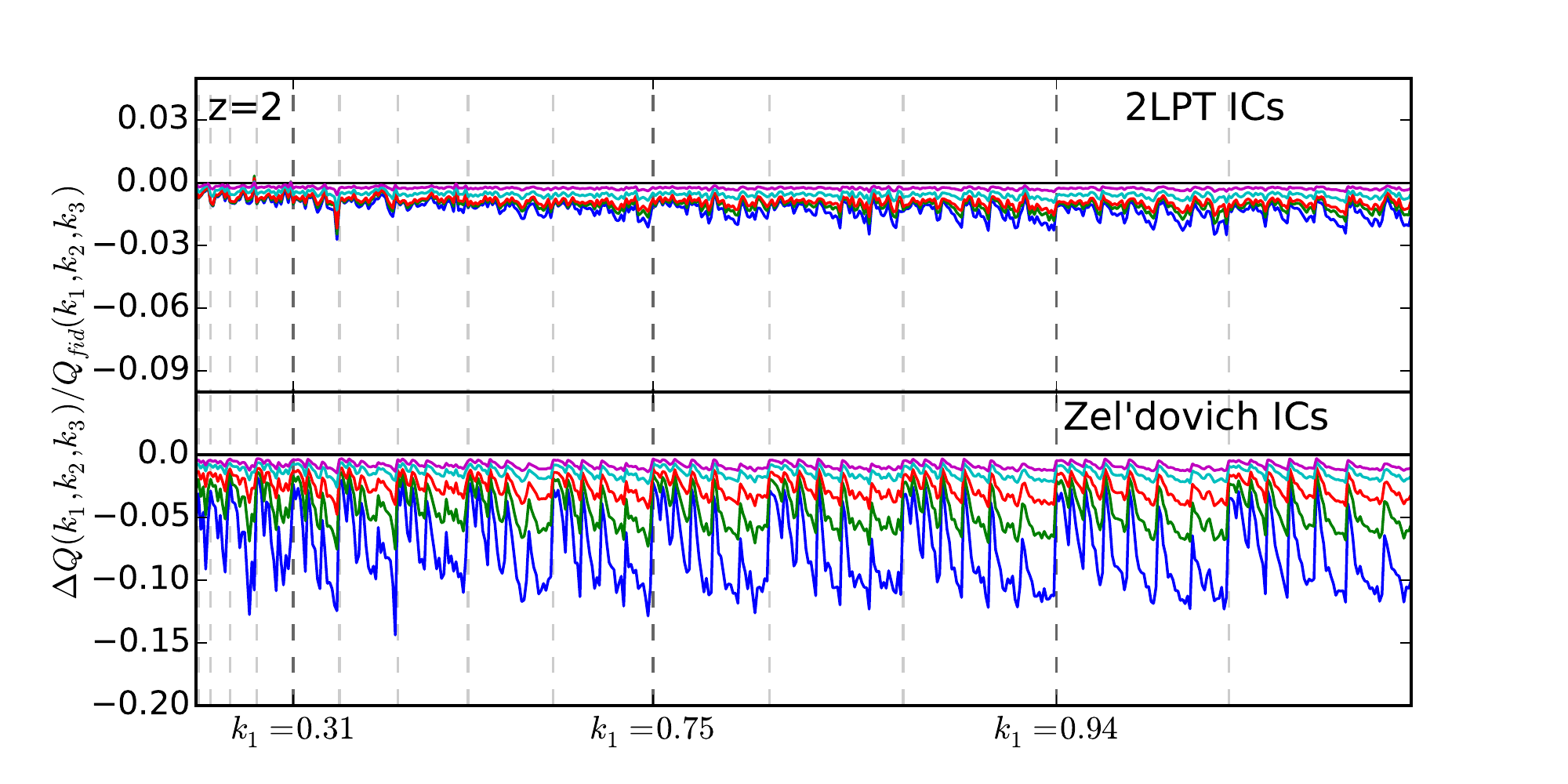}}
\end{subfloat}
\begin{subfloat}[Transient effect at $z=3$]{
\includegraphics[width=0.47\textwidth]{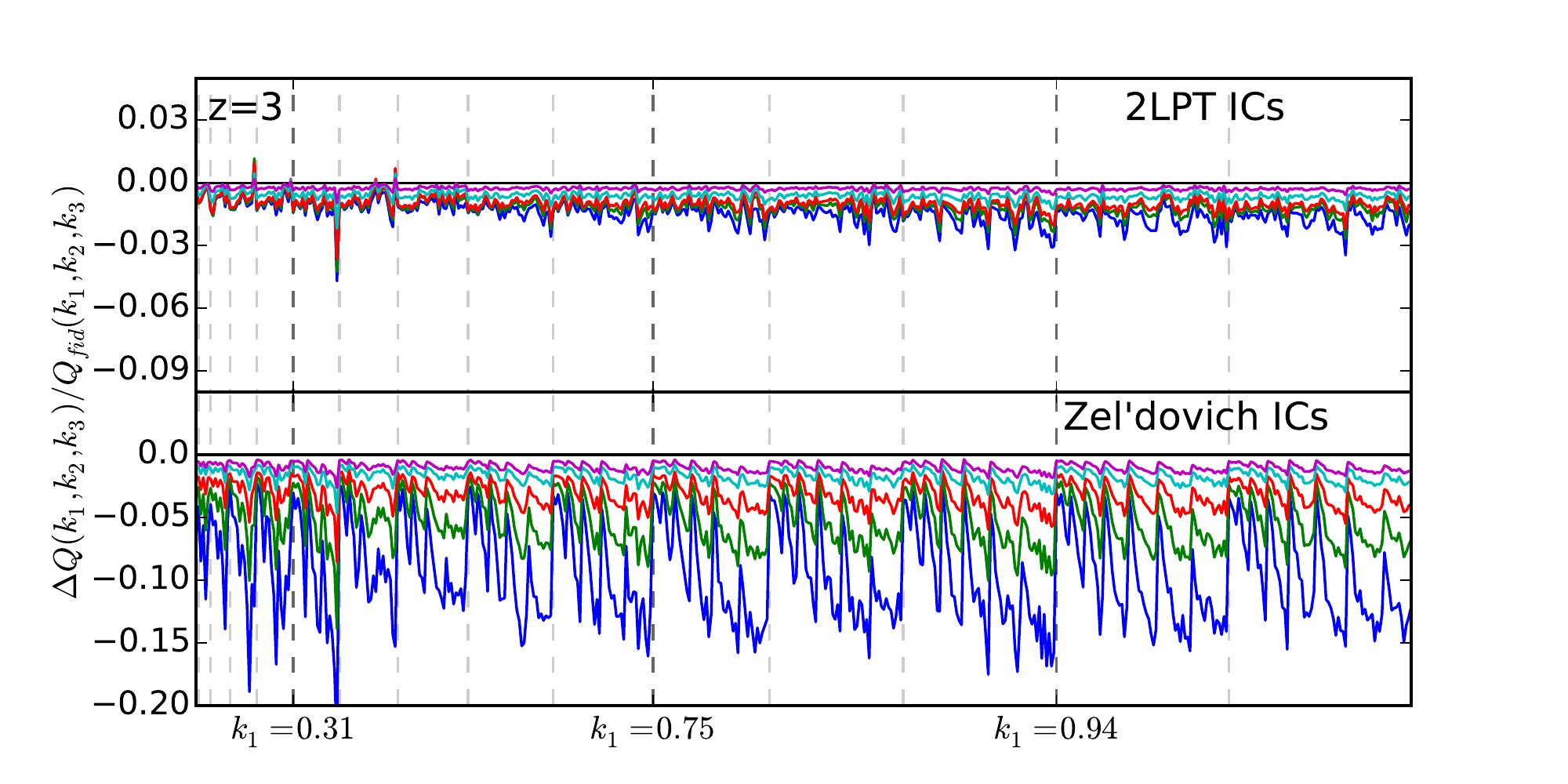}}
\end{subfloat}
\begin{subfloat}[Transient effect at $z=6$]{
\includegraphics[width=0.47\textwidth]{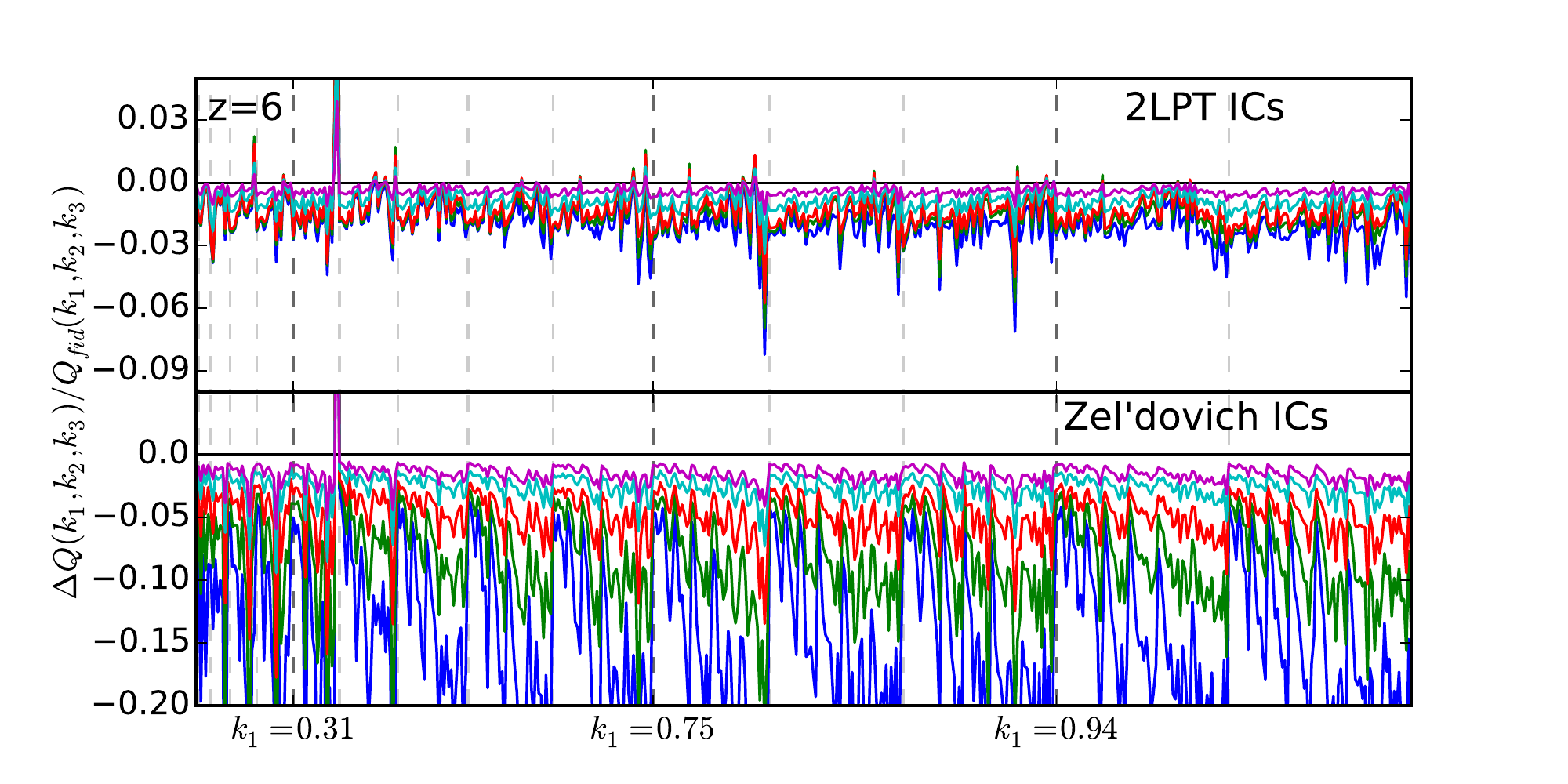}}
\end{subfloat}
\caption{Same as Figure \ref{fig:btransall}, but for the reduced bispectrum, $Q$.}
\label{fig:qtransall}
\end{figure*}

Figure \ref{fig:btransall} shows the fractional error in the measured bispectrum compared to the fiducial bispectrum, again at $z=1, 2, 3$, and $6$, from 2LPT (top) and ZA (bottom) initial conditions. Note the different $y$-ranges in the two panels. As we found with the power spectrum transients, the error is larger for smaller initial redshifts. The bispectrum transient error from the ZA simulations is significantly larger on all scales than in the 2LPT simulations, because the ZA transients appear at tree-level as opposed to 1-loop in PT. We also note that the transient error is larger at higher redshift in all cases, as expected from theory.
We also show the transient effect in the corresponding reduced bispectrum 
\be
Q(k_1,k_2,k_3) = \frac{B(k_1,k_2,k_3)}{P(k_1)P(k_2)+P(k_2)P(k_3)+P(k_3)P(k_1)}
\ee
in Figure \ref{fig:qtransall}. As expected, when removing the effects due to the transients in the power spectrum, the transient effects in 
the reduced bispectrum are smaller, 
in particular for lower redshifts and smaller scales, than the transients in
the bispectrum. The transients in 2LPT simulations, however, are much smaller
than those in ZA simulations.

In order to accurately model the nonlinear bispectrum numerically, we suggest using 2LPT initial conditions with an initial redshift $z_{\rm init} \ge 100$. For $z_i=100$, the error from transients is less than 1\% for $z\le3$ and around $2$\% at $z=6$ for this box size and resolution. For the corresponding ZA simulation, the errors are around 8\% for $z \le 3$ and 10\% for $z=6$.
Although the degree of the transient effect depends on other factors such as the 
resolutions of simulation, we find this rule of thumb also works for 
simulations with $L_{\rm box}= 1~{\rm Gpc}/h$ and $N_{\rm particle}=512^3$.

\begin{figure*}
\centering
\includegraphics[width=0.8\textwidth]{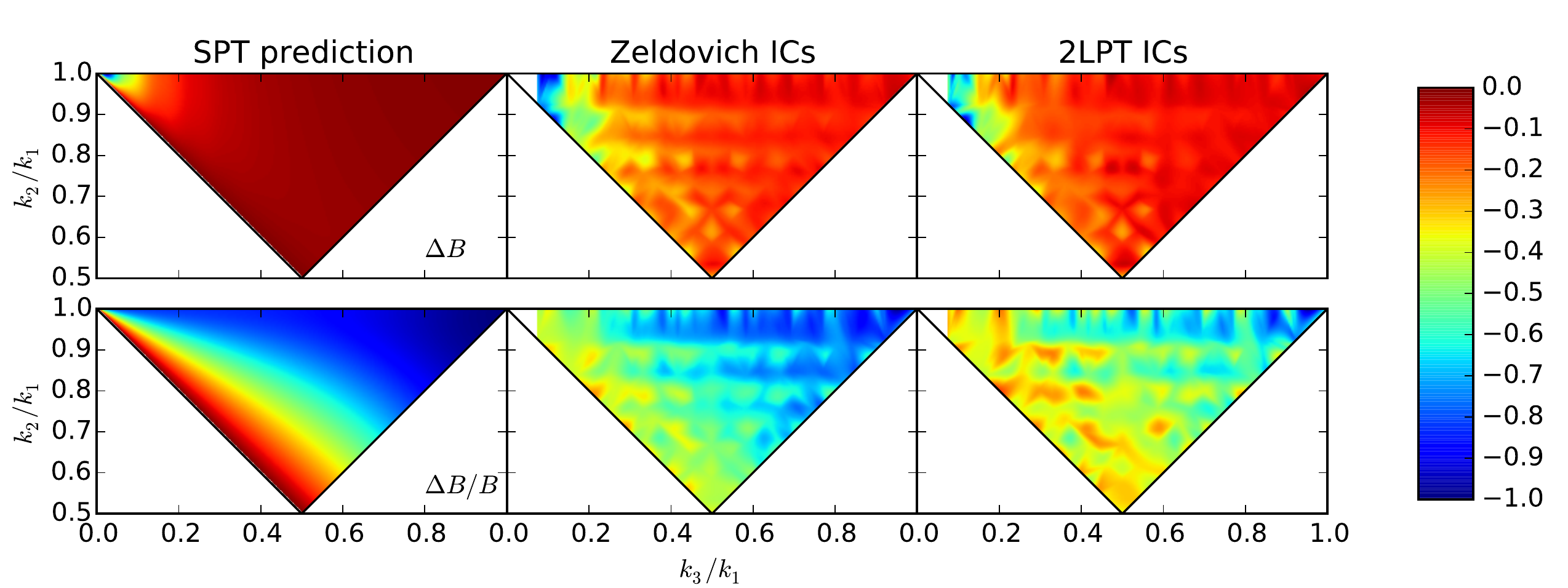}
\caption[Shape dependence of the transient signal and error in the measured bispectrum at $k_1=0.7$ $h$/Mpc at $z=1$ from Zel'dovich and 2LPT initial conditions compared to the SPT prediction.]{Shape dependence of the transient signal and error in the measured bispectrum at $k_1=0.7$ $h$/Mpc at $z=1$ from Zel'dovich and 2LPT initial conditions compared to the SPT prediction. The top panels show the transient signal, which is highest in the squeezed and elongated configurations. The bottom panels show the percent error induced by transients, which is highest in the equilateral configuration. Each slice is normalized by its minimum value, so the range is from $-1$ to $0$ in each case.}
\label{fig:trans_slice}
\end{figure*}

Another consideration is the shape-dependence of the bispectrum transients. Figure \ref{fig:trans_slice} shows the transient signal ($\Delta B=B-B^{\mathrm{fid}}$; top) and induced error ($\Delta B / B^{\mathrm{fid}}$; bottom) at $k_1=0.7$ $h$/Mpc (normalized to $-1$). On the left is the prediction from PT for the transients from ZA initial conditions. The middle column shows the signal and error measured from the the $z_i=200$ ZA simulation at $z=1$. The right column shows the same for the 2LPT simulation with $z_i=200$.

While Figure \ref{fig:btransall} clearly shows that the 2LPT transients have a much smaller amplitude than the ZA transients, Figure \ref{fig:trans_slice} shows that they have very similar shape-dependence to ZA transients on these scales. The measured shape dependences of the transient signal (top panels) and error (bottom panels) agree reasonably well with the SPT prediction for both ZA and 2LPT ICs. The error induced from transients is greatest in the equilateral configuration, and smallest in the elongated configuration for both ZA and 2LPT initial conditions. 
We find that the shape dependence of the transients is 
independent of redshift in both ZA and 2LPT cases;
it is only the amplitude of transients that is reduced at lower redhsifts 
because the decaying modes fade out as the simulation proceeds.

\section{Modeling Nonlinearity in the matter bispectrum}\label{sec:NLmodeling}
We now test the accuracy of various theoretical templates for the 
nonlinear bispectrum against the measured matter bispectrum from 
$N$-body simulations. For comparison, we use the result of 
$z_{\rm init}=400$ (2LPT) simulation, as the nonlinear matter 
bispectra measured from other 2LPT simulations converge to it.

On large enough scales, the leading order (tree level) prediction in 
\refeq{Bktree} is enough to model the matter bispectrum.
Modeling the matter bispectrum on smaller scales where nonlinearities
become stronger, however, requires more elaborate calculation.
On quasi-linear scales, we expect the next-to-leading order (one-loop) bispectrum in PT to model the nonlinear evolution reasonably well.
The next-to-leading order matter bispectrum is given by four 
additional one-loop terms \citep{bernardeau2002}:
\ba
B^{\rm 1loop}(k_1,k_2,k_3)
=&
B^{\rm tree}(k_1,k_2,k_3)
+
B^{(222)}(k_1,k_2,k_3)
\vs
+&
B^{(123a)}(k_1,k_2,k_3)
+
B^{(123b)}(k_1,k_2,k_3)
\vs
+&
B^{(114)}(k_1,k_2,k_3),
\label{eq:Bkloop}
\ea
where $B^{\rm tree}$ is the leading order bispectrum in \refeq{Bktree} and 
\ba
&B^{(222)}(k_1,k_2,k_3)
=
8\int \frac{d^3q}{(2\pi)^3}
F_2^{(s)}(\bm{q},\bm{k}_1-\bm{q})
\vs\quad&\times
F_2^{(s)}(-\bm{q},\bm{k}_2+\bm{q})
F_2^{(s)}(\bm{q}-\bm{k}_1,-\bm{k}_2-\bm{q})
\vs\quad&\times
P_L(q)
P_L(|\bm{k}_1-\bm{q}|)
P_L(|\bm{k}_2+\bm{q}|)\\
&B^{(114)}(k_1,k_2,k_3)
=
12
P_L(k_2)P_L(k_3)
\int \frac{d^3q}{(2\pi)^3}
P_L(q)
\vs\quad&\times
F_4^{(s)}(
\bm{q},-\bm{q},
-\bm{k}_2,-\bm{k}_3
)
+(2~\mathrm{cyc.})
\\
&B^{(123a)}(k_1,k_2,k_3)
=6
P_L(k_1)
\int \frac{d^3 q}{(2\pi)^3}
P_L(q)
P_L(|\bm{k}_2-\bm{q}|)
\vs\quad&\times
F_2^{(s)}(\bm{q},\bm{k}_2-\bm{q})
F_3^{(s)}(-\bm{q},\bm{q}-\bm{k}_2,-\bm{k}_1)
+(5~\mathrm{cyc.})
\ea
\ba
&B^{(123b)}(k_1,k_2,k_3)
=6
P_L(k_1)P_L(k_3)
F_2^{(s)}(\bm{k}_1,\bm{k}_3)
\vs\quad&\times
\int \frac{d^3 q}{(2\pi)^3}
F_3^{(s)}(\bm{q},-\bm{q},\bm{k}_3)P_L(q)
+(5~\mathrm{cyc.}),
\ea
are non-linear corrections with $F_n^{(s)}$ being symmetric kernels
encoding the fastest growing mode of the nonlinear density 
evolution (see, e.g. \refsec{theory}). Note that the expression above is
completely determined by the given linear matter power
spectrum $P_L(k)$.

On smaller scales where PT breaks down due to strong nonlinearities, 
various phenomenological models have been proposed as alternative ways of 
predicting the nonlinear behavior of the matter bispectrum.
Here, we consider three such models (PT+, GM model \cite{hgm2012}, and 
SC model \cite{sc2001}, see below for details) for the nonlinear matter bispectrum.

The simplest model that we consider is PT+, where we attribute the entirety of the 
nonlinear evolution of the matter bispectrum to the nonlinear matter power 
spectrum. This leads to an expression that is identical to \refeq{Bktree}, 
but uses the measured (fully-nonlinear) power spectrum from the $N$-body simulation
in place of the linear power spectrum:
\begin{align}
B(k_1, k_2, k_3)&=2F_2^{(s)}(\kk_1, \kk_2) P(k_1) P(k_2)+\text{(2 cyc.)}
\label{eq:PT+}
\end{align}
This model is motivated from the expression of the one-loop bispectrum 
in \refeq{Bkloop}, which can be written as 
$B^{\rm 1loop}(k_1,k_2,k_3) = 2 F_2^{(s)}(\kk_1,\kk_2)P_{\rm 1loop}(k_1)
P_{\rm 1loop}(k_2) + 2\mathrm{(cyc.)} + $(irreducible terms),
with the one-loop power spectrum $P_{\rm 1loop}(k)$.
In the language of renormalized perturbation theory 
\cite{croccescoccimarro2006}, PT+ corresponds 
to fixing the `vertex' (the kernel $F_2^{(s)}$) while using the renormalized 
propagator (power spectrum $P(k)$).

PT+ is a simplified version of the fitting formula proposed by \citet{sc2001} 
(hereafter, SC), where the $F_2^{(s)}$ kernel in \refeq{PT+} is replaced by
the \textit{effective} kernel $F_2^{\rm eff}$ as follows.
\ba
F_2^{\text{\rm eff}}(\kk_i,\kk_j)&=\frac{5}{7} a(n_i, k_i) a(n_j, k_j)\vs
&\qquad+\frac{1}{2} \cos (\theta_{ij})\left(\frac{k_i}{k_j}+\frac{k_j}{k_i}\right)b(n_i, k_i)b(n_j, k_j)\notag\\
&\qquad+\frac{2}{7}\cos^2(\theta_{ij})c(n_i, k_i) c(n_j, k_j).
\ea
Here, $\theta_{ij}$ is the angle between two wavevectors $\kk_i$ and $\kk_j$,
and $F_2^{\rm eff}$ reduces to the PT kernel in \refeq{F2s} when $a=b=c=1$.
The three functions $a(n,k)$, $b(n,k)$, $c(n,k)$ are given as functions of 
the wavenumber $k$ and the effective slope
of the linear power spectrum $n\equiv d\ln P_L(k)/d\ln k$ as 
\ba
a(n,k)&=\frac{  1+\sigma_8^{a_6}(z) [0.7 Q_3(n)]^{1/2}(qa_1)^{n+a_2}   }{ 1+(qa_1)^{n+a_2}   }\\
b(n,k)&=\frac{  1+0.2a_3(n+3)q^{n+3} }{ 1+q^{n+3.5}   }\\
c(n,k)&=\frac{  1+4.5a_4/[1.5+(n+3)^4](qa_5)^{n+3}   }{ 1+(qa_5)^{n+3.5}},
\ea
where $q\equiv k/k_{\rm nl}$ with nonlinear scale $k_{\rm nl}$ defined by
the wavenumber at which the dimensionless power spectrum
$\Delta^2(k)\equiv k^3P_L(k)/(2\pi^2)$ becomes unity 
$\Delta^2(k_\mathrm{nl})\equiv 1$, and varies with redshift as $k_{\mathrm{nl}}\propto (1+z)$. $Q_3(n)\equiv (4-2^n)/(1+2^{n+1})$.
SC fit the parameters $a_1$ through $a_6$ using the bispectrum measured 
from $\mathrm{P^3M}$ simulations \cite{jenkins/etal:1998} 
with $256^3$ particles in a $240~\mathrm{Mpc}/h$ box to find that 
$a_1=0.25$, $a_2=3.5$, $a_3=2$, $a_4=1$, $a_5=2$, $a_6=-0.2$. 

Recently, \citet{hgm2012} (hereafter, GM) proposed extending the SC model 
further with $3$ additional parameters, where $a(n,k)$, $b(n,k)$ and $c(n,k)$ 
in the effective kernel $F_2^{\rm eff}$ are replaced by tilded functions as 
\begin{align}
\tilde a(n,k)&=\frac{  1+\sigma_8^{a_6}(z) [0.7 Q_3(n)]^{1/2}(qa_1)^{n+a_2}   }{ 1+(qa_1)^{n+a_2}   }\\
\tilde b(n,k)&=\frac{  1+0.2a_3(n+3)q^{n+3}(q a_7)^{n+3+a_8} }{ 1+(qa_7)^{n+3.5+a_8}   }\\
\tilde c(n,k)&=\frac{  1+4.5a_4/[1.5+(n+3)^4](qa_5)^{n+3+a_9}   }{ 1+(qa_5)^{n+3.5+a_9}   }.
\end{align}
The definitions for $n$, $q$, $k_{\text{nl}}$, and $Q_3$ are the same as above,
except in this model $n(k)$ is smoothed so the oscillatory features from the 
Baryon Acoustic Oscillations (BAO) are removed (further discussion of this can be found in \cite{hgm2012}). 
Then, GM fit the parameters $a_1$ to $a_9$ using the measured matter bispectrum
from two sets of $N$-body simulations, one with $L=2.4$ Gpc/$h$ 
($768^3$ particles), 
and one with $L_b=1.875$ Gpc/$h$ ($1024^3$ particles),
to find the best-fit parameters of
$a_1=0.484$, $a_2=3.740$, $a_3=-0.849$, $a_4=0.392$, $a_5=1.013$, $a_6=-0.575$, $a_7=0.128$, $a_8=-0.722$, $a_9=-0.926$.

Note that the conditions on which the fitting formulae are developed are 
quite different. While SC have used the matter bispectrum at all triangular
configurations (a total of a million triangles for $k\lesssim 2.3$ $h$/Mpc)
at redshift $z=0$ and $z=1$, GM have only used certain 
configurations ($\theta_{12}/\pi=0.1$, $0.2$, $\cdots$, $0.9$ 
and $k_2/k_1=1.0$, $1.5$, $2.0$, $2.5$, for $k_2<0.4$ Mpc/$h$) 
at four different redshifts $z=0$, $0.5$, $1$, $1.5$.
Moreover, they are subject to different levels of transient errors 
as the $\Lambda$CDM simulations used by SC are generated at $z_{\rm init}=30$  
with the Zel'dovich approximation, and the simulations used by GM are generated
by 2LPT at $z_{\rm init} = 19$ ($2.4$ Gpc/$h$ box) and $z_{\rm init} = 49$
($1.875$ Gpc/$h$ box).

\begin{figure}
\centering
\includegraphics[width=0.5\textwidth]{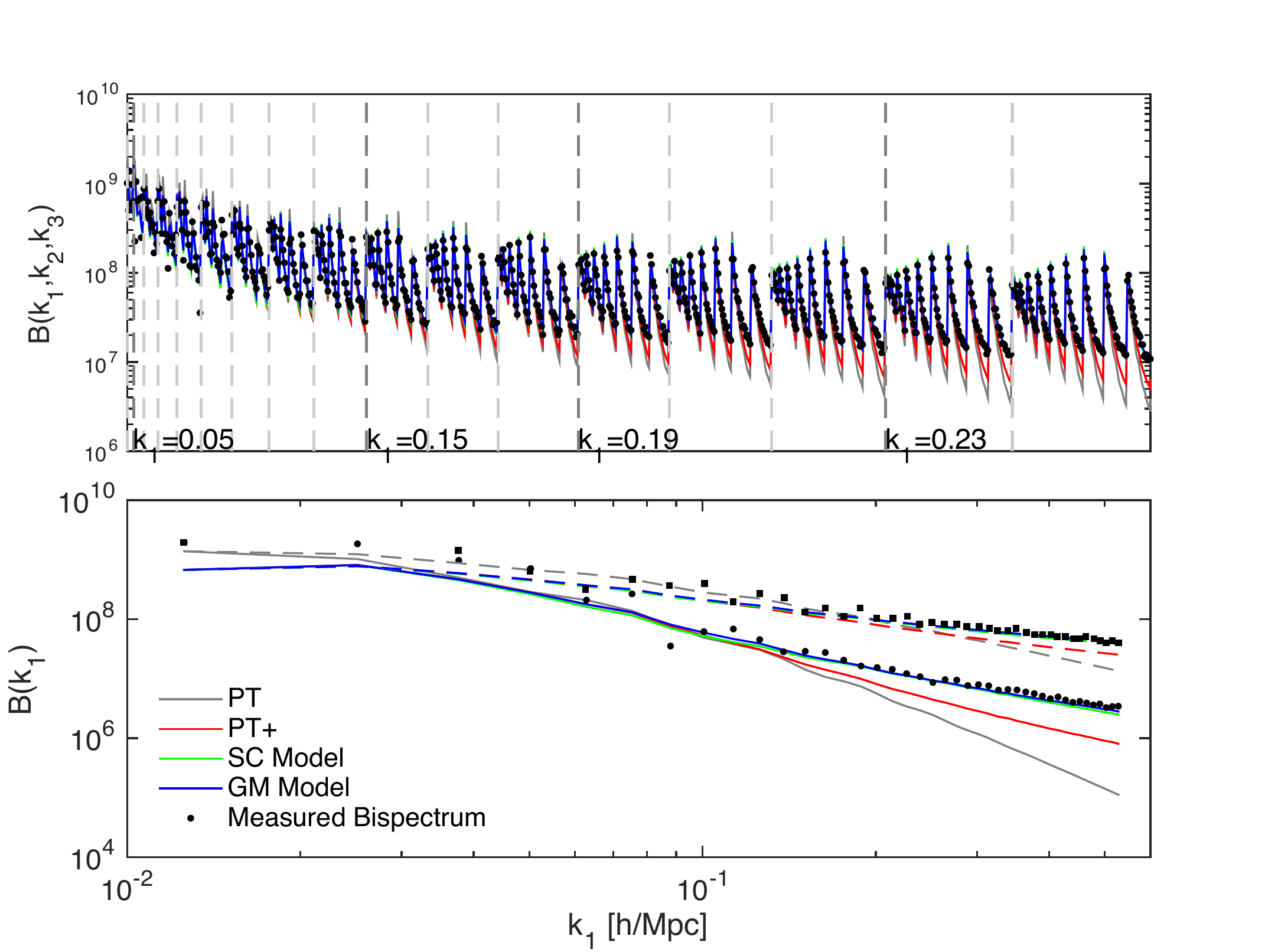}
\caption[The bispectrum measured from a 1 Gpc/$h$ simulation at $z=0$ compared to analytic models of the nonlinear bispectrum.]{The bispectrum at $z=0$ measured from the 1 Gpc/$h$ simulation, compared to the SC (green), GM (blue), PT (grey), and PT+ (red) models. The top plot shows the full flattened bispectrum, and the bottom plot shows only the squeezed and equilateral triangles as a function of $k_1$.}
\label{fig:modelsz0}
\end{figure}

To test the validity of each of the phenomenological formulae outside of the 
dynamical (Fourier) ranges and redshift ranges at which the models are fitted
for, we compare the model predictions to measured bispectra 
at redshifts $1<z<6$ in both large-scale ($L_{\text{box}}=1$ Gpc/$h$) and 
small-scale ($L_{\text{box}}=200$ Mpc/$h$) simulations. We also compare these to the predictions from tree-level PT, PT+, and 1-loop PT. 
For both the SC and GM models, we calculate the effective slope of the linear
power spectrum $n(k)$ from the BAO-smoothed linear power spectrum which 
provides better agreement than the $n(k)$ including BAO.

\begin{figure*}
\centering
\includegraphics[width=1.0\textwidth]{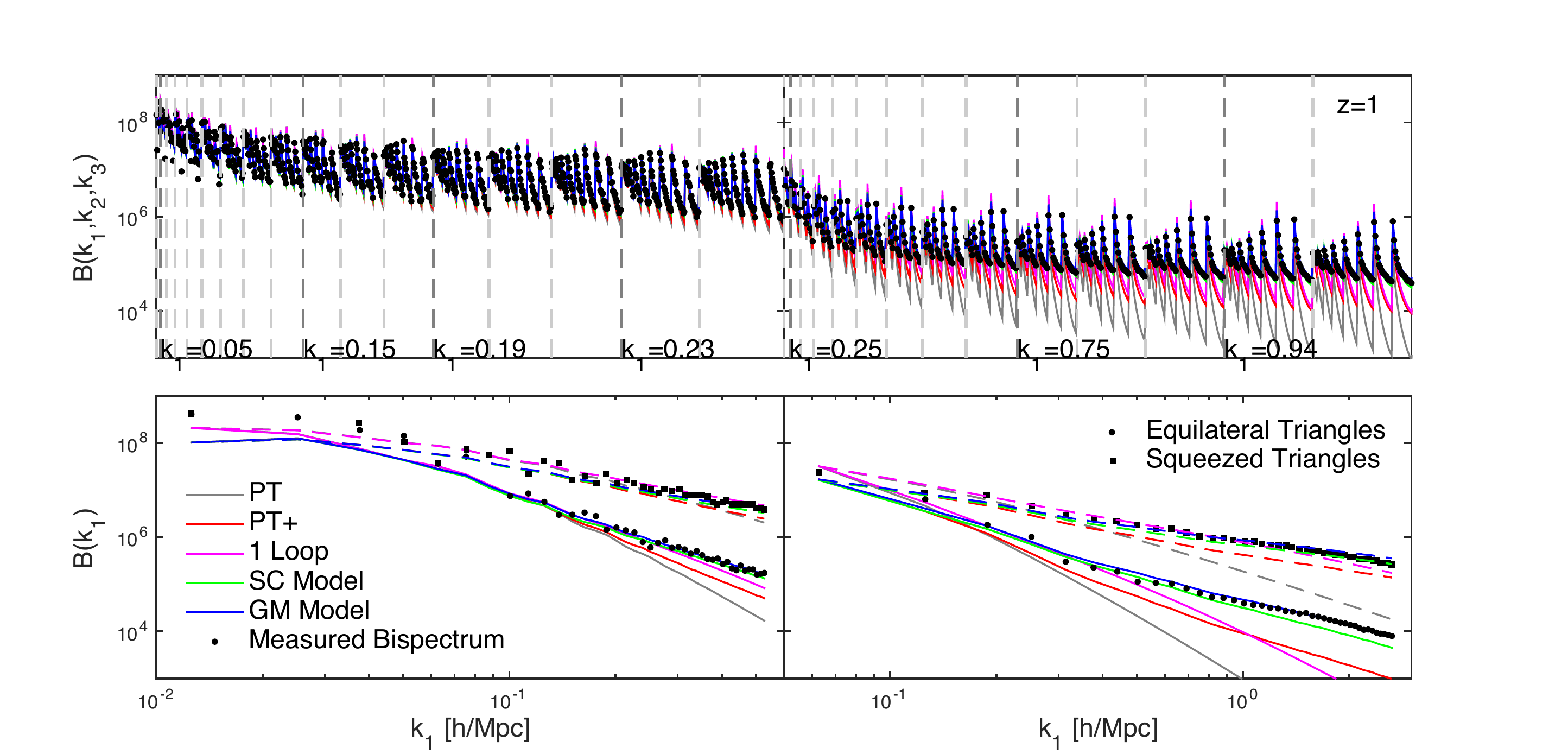}
\caption[The bispectrum measured from a 1 Gpc/$h$ and a 200 Mpc/$h$ simulation at $z=1$ compared to analytic models of the nonlinear bispectrum.]{The bispectrum at $z=1$ measured from the large-box (left) and small-box (right) simulations, compared to the SC (green), GM (blue), PT (grey), and PT+ (red) models. The top plot shows the full flattened bispectrum, and the bottom plot shows only the squeezed and equilateral triangles as a function of $k_1$.}
\label{fig:modelsz1}
\end{figure*}

First, we check that we can reproduce the results from \citet{hgm2012} at $z=0$ on large scales. We expect that the GM model is the most accurate here, because their model parameters were fit from a similar simulation. Figure \ref{fig:modelsz0} shows the measured bispectrum at $z=0$ from the 1 Gpc/$h$ simulation, compared to the SC and GM models, as well as PT and PT+. The top plot shows the full flattened bispectrum, and the bottom plot shows only the squeezed and equilateral triangles as a function of $k_1$. We can see that the GM model indeed gives an improved agreement to the measured bispectrum over the SC model on these scales at $z=0$.

\begin{figure*}
\centering
\includegraphics[width=1.0\textwidth]{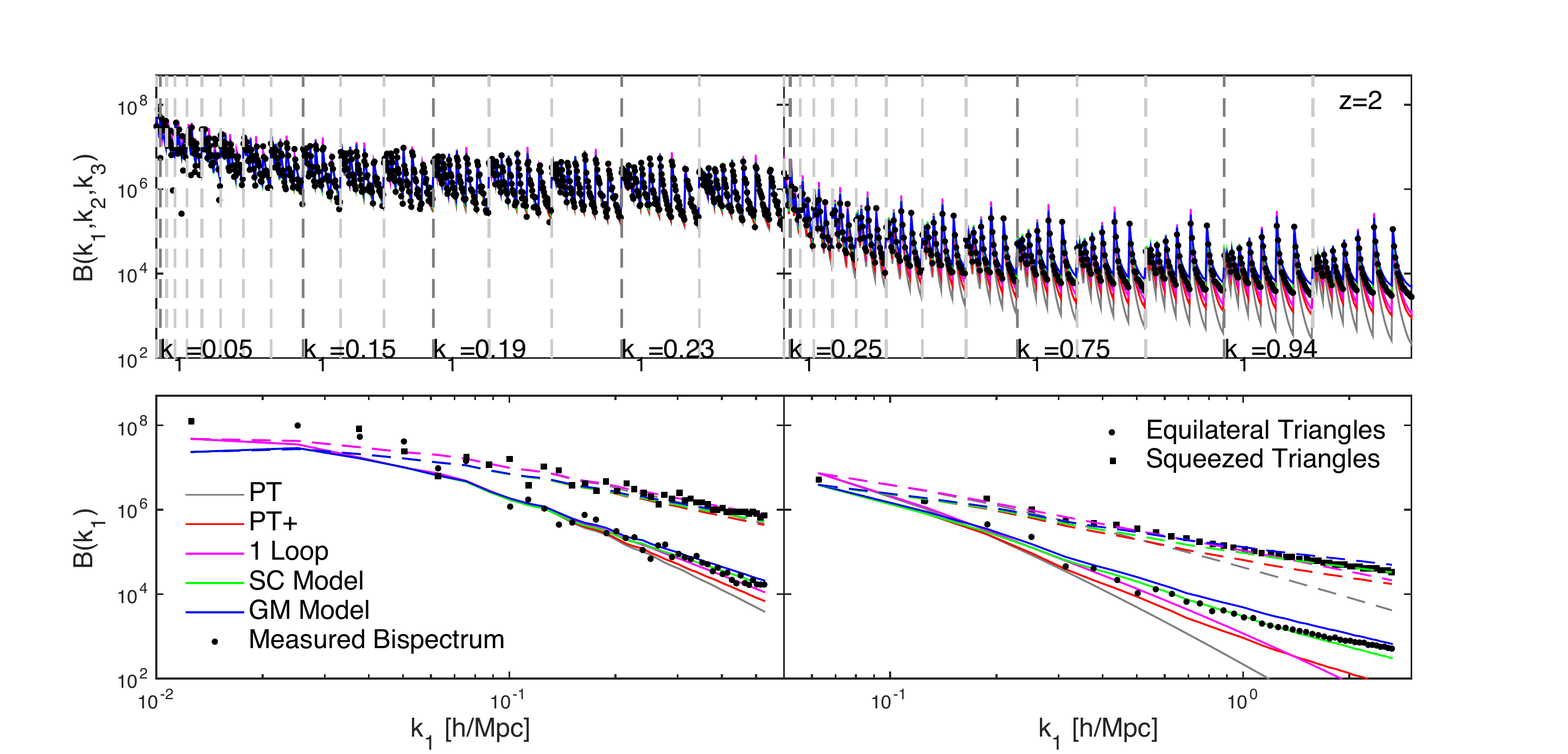}
\caption[The bispectrum measured from a 1 Gpc/$h$ and a 200 Mpc/$h$ simulation at $z=2$ compared to analytic models of the nonlinear bispectrum.]{Same as Fig. \ref{fig:modelsz1} but for $z=2$. }
\label{fig:modelsz2}
\end{figure*}

Next we look at $z=1$ in both the 1 Gpc/$h$ and 200 Mpc/$h$ simulations. Figure \ref{fig:modelsz1} compares the measured bispectrum to the models at $z=1$. The left column shows the results from the large-box simulation, and the right shows the small-box simulation. Note that the $k_1$ range in the bottom plots overlaps, but we plot the results from the two simulations separately for clarity. Also note that the measured signal for squeezed triangles from the two simulations is slightly different due to the different resolutions. From these plots, we see that the GM model still gives the best fit to the measured bispectrum on the scales shown in all triangular configurations, except for squeezed triangles on very small scales, where the SC model fits slightly better.

\begin{figure*}
\centering
\includegraphics[width=1.0\textwidth]{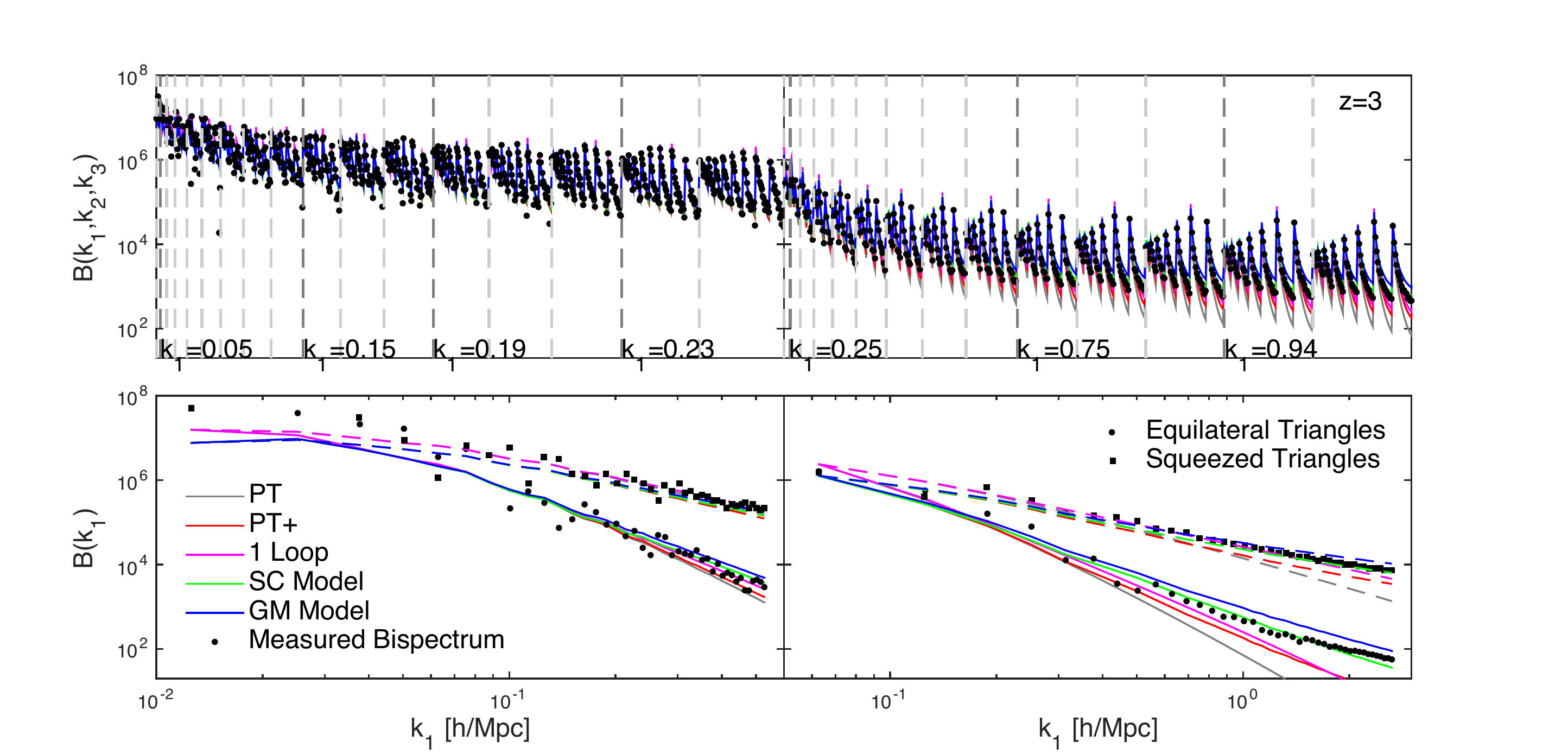}
\caption[The bispectrum measured from a 1 Gpc/$h$ and a 200 Mpc/$h$ simulation at $z=3$ compared to analytic models of the nonlinear bispectrum.]{Same as Fig. \ref{fig:modelsz1} but for $z=3$. }
\label{fig:modelsz3}
\end{figure*}

\begin{figure*}
\centering
\includegraphics[width=1.0\textwidth]{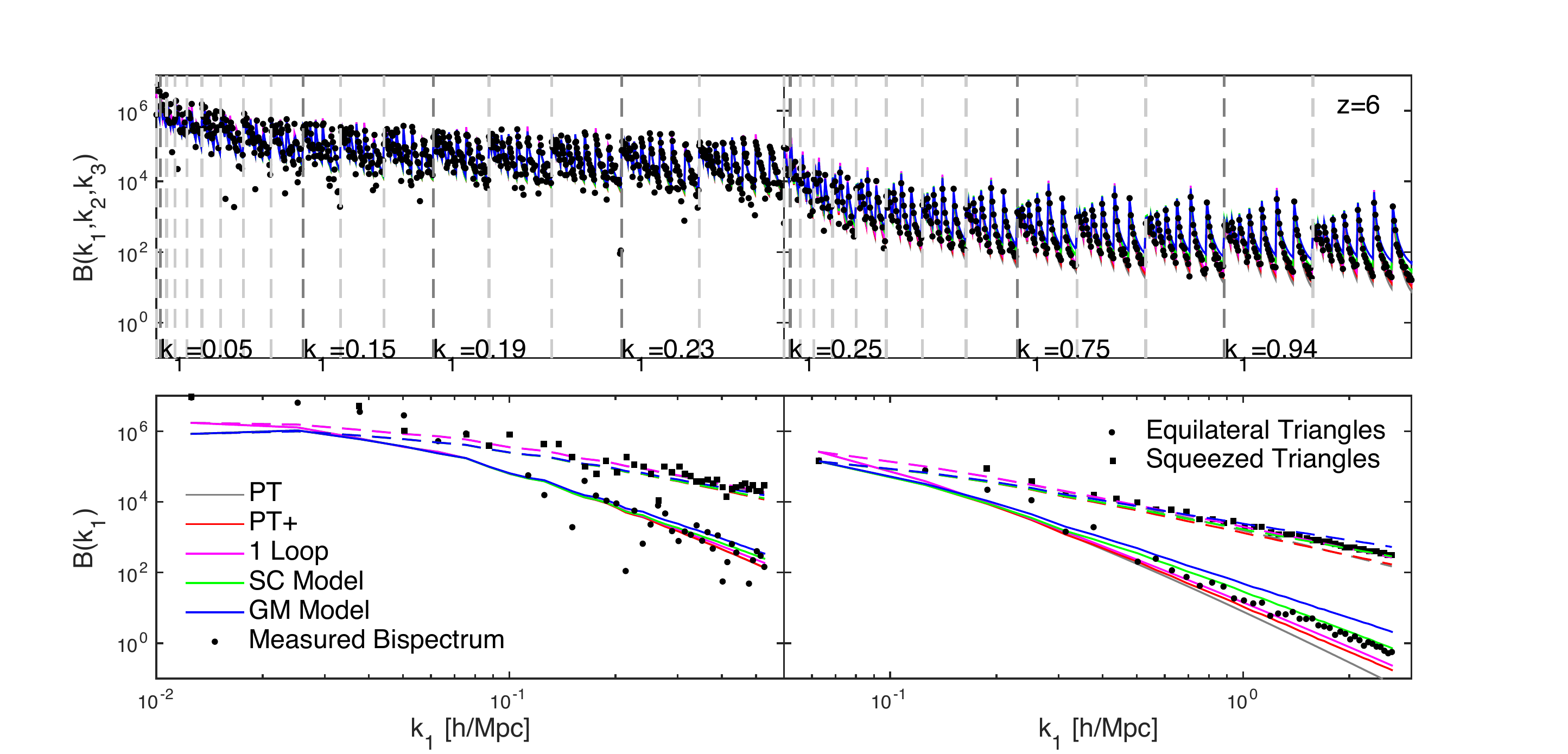}
\caption[The bispectrum measured from a 1 Gpc/$h$ and a 200 Mpc/$h$ simulation at $z=6$ compared to analytic models of the nonlinear bispectrum.]{Same as Fig. \ref{fig:modelsz1} but for $z=6$. }
\label{fig:modelsz6}
\end{figure*}

However, we find at higher redshifts, the GM model is less accurate than the SC model. Figs. \ref{fig:modelsz2}, \ref{fig:modelsz3}, and \ref{fig:modelsz6} show the measured bispectrum at $z=2$, $3$, and $6$, respectively, from both simulations compared to the models. In all of these cases, the SC model gives the most accurate prediction on all scales and in all triangular configurations. The GM model over-predicts the bispectrum slightly in all cases. On large scales (left panels), the predictions from perturbation theory (1-loop and PT+) give reasonably accurate predictions.

It is clear from these comparisons that one must carefully consider the range of validity of a given model when predicting the nonlinear bispectrum. At $z\le1$, the GM model undoubtedly gives the best prediction, but at higher redshifts, it is better to use the SC model. At high redshift ($z\ge 3$), the PT models are sufficient for modeling the nonlinear bispectrum. Further work is needed to develop a more reliable model of the nonlinear bispectrum on all scales and redshifts.

\section{Conclusion}\label{sec:conclusion}

As the non-linear growth of structure generally causes the cosmic density 
field to become non-Gaussian, studying higher-order statistics beyond the two-point 
correlation function will open up a new avenue to exploit a wealth of 
cosmological information. In particular, with a large-volume coverage and 
high number density of galaxies, current and future galaxy surveys promise 
ever-more accurate measurements of the galaxy three-point function and bispectrum. These measurements of three-point correlation functions 
can give us a better understanding of the growth of structure, galaxy bias, 
and primordial non-Gaussianity. 

Extracting information from the galaxy bispectrum requires accurate modeling 
of the bispectrum of the dark-matter density field. We have shown that when 
studying the matter bispectrum numerically, it is important to understand the 
effects of transients from initial conditions in simulations. When 
generating the initial conditions using Zel'dovich approximation, the 
transient of the matter bispectrum shows up at leading order in 
perturbation theory and affects all scales.
In contrast, the transient of the matter power spectrum shows up only at higher-order and affects mainly small scales.
In order to simulate nonlinearity in the matter bispectrum correctly,
a 2LPT (second-order Lagrangian perturbation theory) initial conditions
generator must be used. To further reduce the effects of higher-order transients, 
we recommend using 2LPT initial conditions with $z_{\text{init}}>100$.
This requirement is based on our suite of simulations, in order to maintain
a sub-percent level accuracy of the matter bispectrum on scales $k<1\ h$/Mpc 
at $z\le3$.

Another challenge is theoretical modeling of the nonlinear behavior of 
the bispectrum beyond tree-level perturbation theory. 
We have shown that analytical calculations of the next-to-leading order,
 one-loop bispectrum models the non-linearities in the matter bispectrum 
quite well on quasi-nonlinear scales, in particular at high redshifts.
The fact that this analytical calculation of the one-loop bispectrum, which does not contain any
free parameters, can model the non-linearities in the matter bispectrum on
quasi-nonlinear scales is very encouraging for theoretical modeling.
More systematic and in-depth studies, 
parallel to \cite{jeong/komatsu:2006} 
but for the matter bispectrum, are needed in order to assess the accuracy of 
perturbation theory modeling at high redshifts in the quasi-nonlinear regime.

On even smaller, non-linear scales, we found that
the fitting formula from \citet{hgm2012} is reliable at low redshift ($z\le1$), but at higher redshifts, \citet{sc2001} gives the best agreement with the bispectrum measured from simulations. We caution the readers to critically check the range of validity of these formulae before applying them.

\section*{Acknowledgements}
We thank Hector Gil-Mar\'in for useful discussion. DJ thanks Max-Planck Institute for Astrophysics for hospitality.

\bibliographystyle{mnras}
\bibliography{bkiniZ}

\end{document}